\documentclass[runningheads]{llncs}
\usepackage{amsmath}
\usepackage{amsfonts}
\usepackage{graphicx}
\usepackage{subfig}
\usepackage{cite}
\usepackage{hyperref}
\usepackage{float}
\usepackage{booktabs}

\begin{document}

\title{Learning a low dimensional manifold of real cancer tissue with PathologyGAN}
\titlerunning{ }

\author{Adalberto Claudio Quiros\inst{1}\orcidID{0000-0003-4804-0741} \and
Roderick Murray-Smith\inst{1}\orcidID{0000-0003-4228-7962} \and
Ke Yuan\inst{1}\orcidID{0000-0002-2318-1460}}
\authorrunning{Quiros A.C. et al.}

\institute{University of Glasgow, School of Computing Science, UK \\
\email{a.claudio-quiros.1@research.gla.ac.uk, \\ \{roderick.murray-smith, ke.yuan\}@glasgow.ac.uk}}

\maketitle              
%

\begin{abstract}


Application of deep learning in digital pathology shows promise on improving disease diagnosis and understanding. We present a deep generative model that learns to simulate high-fidelity cancer tissue images while mapping the real images onto an interpretable low dimensional latent space. The key to the model is an encoder trained by a previously developed generative adversarial network, PathologyGAN. We study the latent space using 249K images from two breast cancer cohorts. We find that the latent space encodes morphological characteristics of tissues (e.g. patterns of cancer, lymphocytes, and stromal cells). In addition, the latent space reveals distinctly enriched clusters of tissue architectures in the high-risk patient group.

\keywords{Generative Adversarial Networks \and Digital Pathology.}
\end{abstract}

\section{Introduction}
Diagnosis and treatment of cancer are commonly based on assessment of histopathological images such as the haematoxylin and eosin (H\&E) stained tissue images. The clinical utility of H\&E images is due to the rich information about the tumor microenvironment recorded in them, such as the phenotype of cancer cells, immune cells, tissue architecture, and how they interact. Recently,  advanced machine learning and deep learning approaches have been developed to improve our understanding of the tumor microenvironment \cite{Campanella2019}. A common theme of these approaches is to correlate the quantification of tumor microenvironment to a known clinically significant phenotype \cite{beck/systematic_analysis/2011, yuan/quantitative_cellular/2012} and molecular characteristics \cite{Fu2019,Coudray2018}. The quality of such correlation-based studies largely depends on the heterogeneity within the response and explanatory variables. Large cancer genome sequencing projects have revealed substantial diversities of molecular and clinical characteristics within and between patients \cite{Ciriello2013}. Although it has been studied in breast and ovarian cancers \cite{Natrajan2016,Zhang2018}, the heterogeneity of tumor microenvironment is largely unknown.     

Here, we propose a representation learning and disentanglement framework for unsupervised quantification and clustering of tissue architectures, which relates phenotype and patient survival. We use Generative Adversarial Networks (GANs) as a tool to find useful representation of cancer tissue architectures. We summarize our contributions as:
\begin{enumerate}
    \item Based on PathologyGAN model \cite{quiros2019pathology}, we introduce an encoder that can be trained to act as an inverse function of the generator, taking advantage of the generator's ability to capture tissue characteristics. This allows us to project real tissue onto the generative model's latent space.
    \item We demonstrate that the encoder is able to interpret the morphological attributes of the cancer tissue and place the tissue images in distinct regions of the latent space. 
    \item We capture the change of cancer tissue morphology densities between patients with survival times greater and lesser than five years, aligning with with previous findings \cite{beck/systematic_analysis/2011}.
\end{enumerate}

\section{Background}
    Generative Adversarial Networks \cite{DBLP:journals/corr/GoodfellowPMXWOCB14} are models that are able to learn diverse and faithful data representations from a given distribution. This is done with a generator, $G(z)$, that maps random noise, $\boldsymbol{z} \sim p_{\boldsymbol{z}}(z)$, to samples that resemble the target data, $\boldsymbol{x} \sim p_{\text { data }}(\boldsymbol{x})$, and a discriminator, $D(x)$, whose goal is to distinguish between real and generated samples. The goal of a GAN is find the equilibrium in the min-max problem:
    \begin{equation}
        \min _{G} \max _{D} V(D, G)=\mathbb{E}_{\boldsymbol{x} \sim p_{\text { data }}(\boldsymbol{x})}[\log D(\boldsymbol{x})]+\mathbb{E}_{\boldsymbol{z} \sim p_{\boldsymbol{z}}(\boldsymbol{z})}[\log (1-D(G(\boldsymbol{z})))].
    \end{equation}

    GANs have since improved in image resolution, quality, and diversity with models as SNGAN \cite{DBLP:conf/iclr/MiyatoKKY18}, BigGAN \cite{DBLP:journals/corr/abs-1809-11096}, ProGAN \cite{DBLP:conf/iclr/KarrasALL18}, RealsnessGAN \cite{Xiangli*2020Real}, and StyleGAN \cite{Karras2018ASG, karras2019analyzing}. There has been an increased focus on improving GANs for disentanglement and representation learning, such as InfoGAN \cite{chen2016infogan}, BiGAN \cite{donahue2016adversarial}, StyleGAN \cite{Karras2018ASG, karras2019analyzing}, and BiBigGAN \cite{donahue2019large}. These models allow certain control in image generation with specific feature properties. 
    Simultaneously, projecting real images onto a GAN's latent space has also gained interest in the literature. Some works have used pre-trained generators and to find the real image projections through an iterative process \cite{Abdal_2019, lipton2017precise, Karras2018ASG, karras2019analyzing}, yet these methods are usually costly and image-by-image based. Alternatively, other models include an encoder with different optimization goals like gradual latent space \cite{sainburg2018generative}, representation learning \cite{donahue2016adversarial, donahue2019large, makhzani2015adversarial}, or disentanglement \cite{makhzani2015adversarial}. Given the computational capacity and current state of generative models for representation learning, GANs or VAEs can make an impact on real-task applications such as histopathology.
    
    Machine learning and especially deep learning approaches have shown early success in digital pathology, not only on achieving high accuracy classification \cite{Esteva2017,DBLP:journals/corr/abs-1901-11489, sci_reports/breast_cancer_multi/2017}, but also in assisting in the decision process with computer-human interaction \cite{DBLP:conf/chi/CaiRHHKSWVCST19}. These methods are usually either supervised or weakly supervised, which require previous knowledge to train the models. On the other hand, unsupervised models only require the data samples (cancer tissue images) to find common attributes or properties that can explain the data.
    Unsupervised models are gaining interest in histopathology and they have been applied to tasks including tissue or nuclei segmentation \cite{7163353, Hou_2019, 10.1117/12.2293717, 8642295, pmlr-v102-gadermayr19a}, classification \cite{bulten2018unsupervised}, high resolution image generation 
    \cite{Levine2020.02.24.963553, quiros2019pathology}, and representation learning \cite{Hou_2019, quiros2019pathology}). 
    Our work focuses on building a GAN with an encoder that has disentanglement and representation learning properties, and that way providing a framework for unsupervised quantification and clustering of tissue architectures.
    \begin{figure}[t]
        \centering
        \includegraphics[scale=0.17]{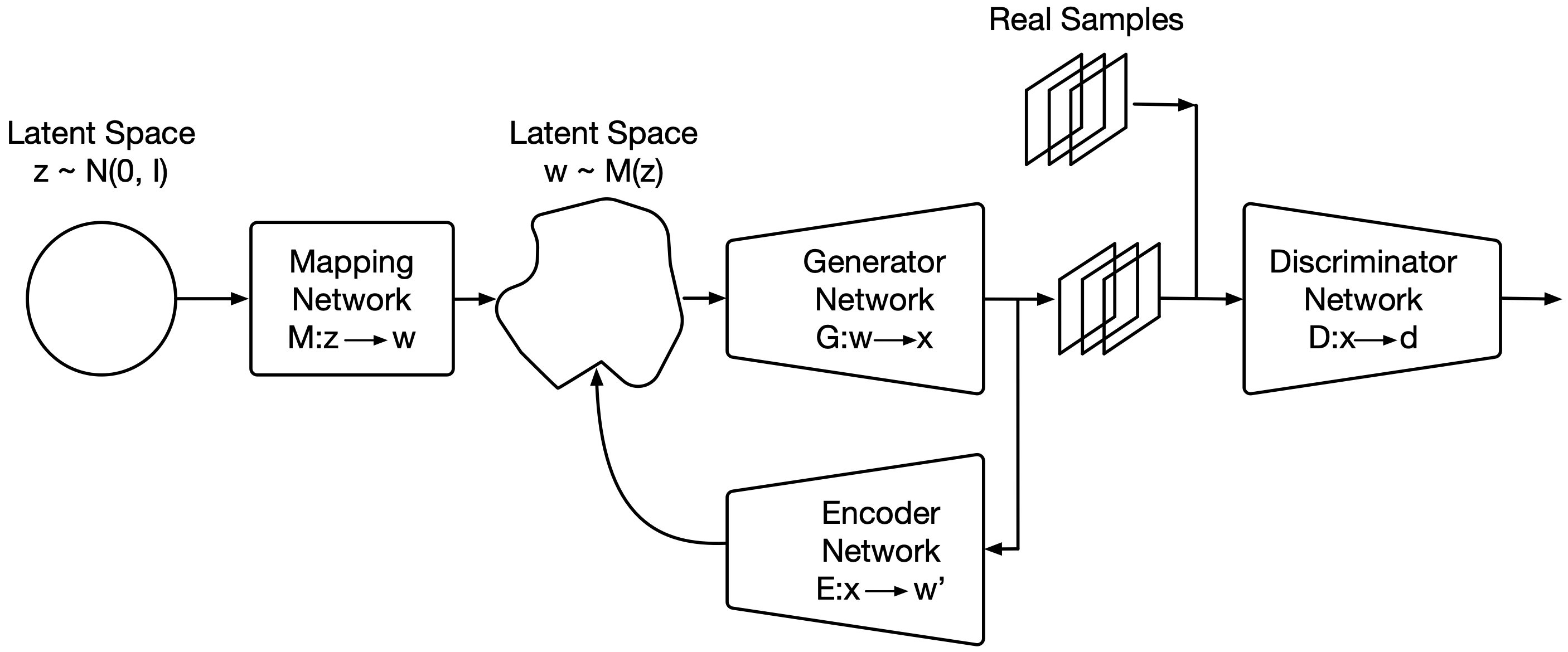}
        \caption{High level architecture of our GAN model.}
        \label{fig:pathologygan_enc_model}
    \end{figure}
    
\section{PathologyGAN Encoder}
    We build upon PathologyGAN \cite{quiros2019pathology}, which used techniques from BigGAN \cite{DBLP:journals/corr/abs-1809-11096} and StyleGAN \cite{Karras2018ASG} to successfully reproduce cancer tissue images while having an interpretable latent space. 
    In our model, the encoder $E$ learns to interpret tissue morphology through generated images, effectively acting as the inverse of the generator $G$. In PathologyGAN, the generator has disentanglement and representation learning properties. We take advantage of this fact by forcing the encoder to learn to place generated images back to the latent space. This process trains an encoder that is able to map tissue with different properties (e.g. cancer cell, lymphocyte, and stromal density) to distinct regions of the latent space. Figure \ref{fig:pathologygan_enc_model} captures the high level network architecture of our model. After training, the encoder can be used independently to map real images to their representations in the latent space. 
    
    We define the loss functions for the discriminator as $L_{Dis}$ and the generator as $L_{Gen}$, which remain the same as in the GAN model (Equations 2 and 3): 
    \begin{align}
        L_{Dis}=-\mathbb{E}_{x_{r} \sim P_{data}}&\left[\log \left(\tilde{D}\left(x_{r}\right)\right)\right]-\mathbb{E}_{x_{f} \sim G(w)}\left[\log \left(1-\tilde{D}\left(x_{f}\right)\right)\right], \\
        L_{Gen}=-\mathbb{E}_{x_{f} \sim G(w)}&  \left[\log\left(\tilde{D}\left(x_{f}\right)\right)\right]-\mathbb{E}_{x_{r} \sim P_{data}}\left[\log \left(1-\tilde{D}\left(x_{r}\right)\right)\right], \\
        \quad \quad \tilde{D}\left(x_{r}\right) &= \text{sigmoid}\left(C\left(x_{r}\right)-\mathbb{E}_{x_{f} \sim G(w)} C\left(x_{f}\right)\right)), \nonumber \\ 
        \quad \quad \tilde{D}\left(x_{f}\right) &= \text{sigmoid}\left(C\left(x_{f}\right)-\mathbb{E}_{x_{r} \sim P_{data}} C\left(x_{r}\right)\right), \nonumber \\
        \quad \quad w &\sim M(z),\ z \sim P_{z}. \nonumber
        \label{eqn:disc_loss}
    \end{align} 
    The encoder loss function, $L_{Enc}$, is defined to minimize the mean square error of the latent vectors $w$ and its reconstruction $w'=E(G(w))$ through generated images (Equation 4):  
    \begin{align}
        L_{Enc} &= \mathbb{E}_{z \sim P_{z}}\left[ \frac{1}{n} \sum_{i=1}^{n}(w_{i} - w'_{i})^{2}\right] \\
        \quad \quad n &= dim(w),\ w'=E(G(w)),\ w \sim M(z). \nonumber
        \label{eqn:disc_loss}
    \end{align} 
    
    Although the encoder $E$ is simultaneously trained with the GAN model, we can separate the model training into two parts: the mapping network $M$, generator $G$, and discriminator $D$ that are trained as a GAN with Relativistic Average Discriminator \cite{DBLP:journals/corr/abs-1807-00734}, and the encoder $E$, which is trained to project back the generated cancer tissue images onto the latent space. In practice, the encoder $E$ learns simultaneously with the Generator $G$. 
    
    We trained our encoder based on the assumption that the generator is successful in reproducing real cancer tissue. Therefore the encoder will learn to project real tissue images if it is able to do so with generated ones. Based on this logic, we use only generated images to train the encoder. 
    
    The encoder is only updated when the generator is not trained with style mixing regularization \cite{Karras2018ASG}. Style mixing regularization uses two latent vectors $w_{1}$ and $w_{2}$ to force disentanglement in the generator. It becomes impractical to train the encoder in these steps because these images have no clear assignation in the latent space. 
    The style mixing regularization is only preformed $50$\% of times in the generator training, so our encoder is updated every two steps per the generator.
    
    To train our model we used two haematoxylin and eosin (H\&E) breast cancer databases from the Netherlands Cancer Institute (NKI) cohort and the Vancouver General Hospital (VGH) cohort with 248 and 328  patients,  respectively \cite{beck/systematic_analysis/2011}. In total, this corresponded to a training set of $249K$ images of $224\times224$ pixels. We used an NVIDIA  Titan  RTX  24  GB to train the model for approximately 80 hours. 
    
    \begin{figure}[!b]
        \centering
        \includegraphics[scale=0.12]{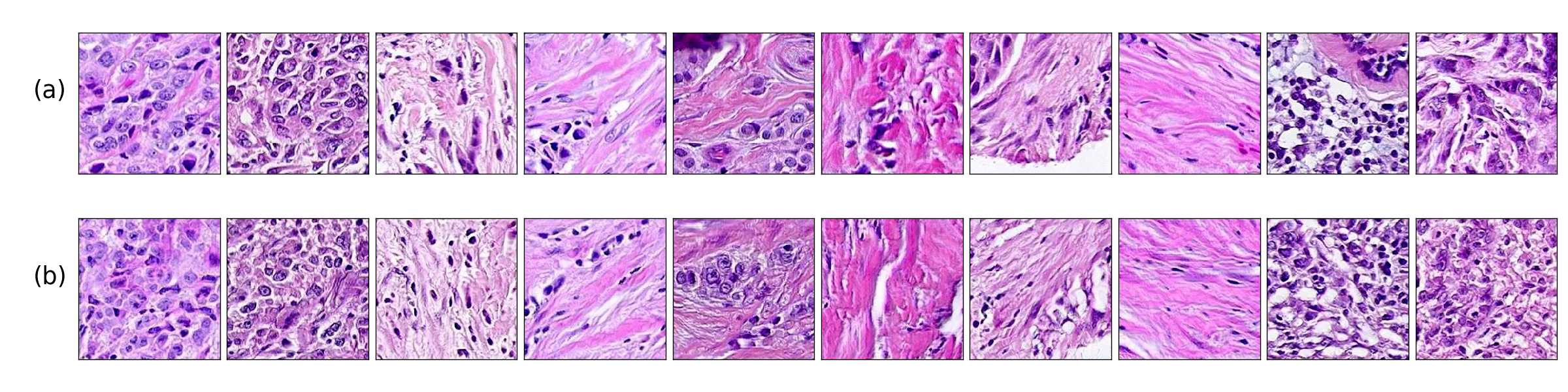}
        \caption{Real tissue images and its reconstructions. We take real tissue images and map them to the latent space with our encoder, then we use the generator with the latent vector representations to generate the image reconstructions. (a) correspond to the real tissue images and (b) to the reconstructions, the images are paired in columns. We show different examples of stromal, lymphocytes, cancer cells, and combinations of these, the reconstructions follow the real image attributes.}
        \label{fig:real_recon}
    \end{figure}
    
    \begin{figure}[!b]
        \centering
        \includegraphics[scale=0.14]{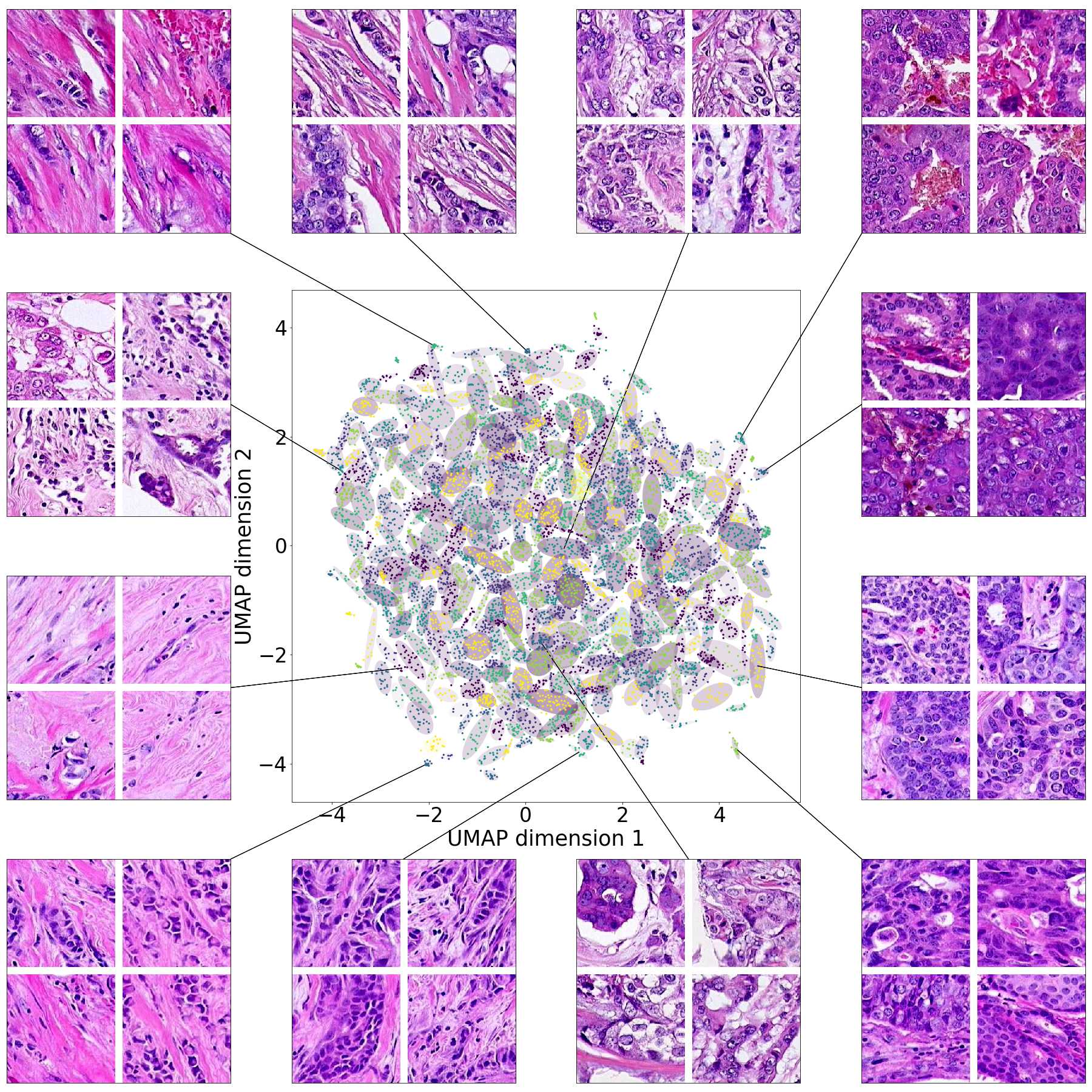}
        \caption{Uniform Manifold Approximation and Projection (UMAP) representation of real tissue samples in the latent space using samples from Netherlands Cancer Institute (NKI) and Vancouver General Hospital (VGH) patient cohorts. In this Figure, we fitted a Gaussian mixture model over the complete dataset and used 100 components to cluster the latent representations. We show different tissue images belonging to various unique clusters, demonstrating how tissues with similar features get assigned to common regions in the latent space.}
        \label{fig:real_img_cluster}
    \end{figure}
    
\section{Results and discussion}
    Our results focus on analyzing our model's comprehension of tissue characteristics, such as colour, texture, spatial features of cancer, lymphocytes, and stromal cells. For these results we used only real H\&E breast tissue samples of VGH and NKI cohorts.
    
\subsection{Tissue image Reconstruction}
    We start by analyzing how much information about the tissue the is model capturing. The assumption is that if the encoder truly finds meaningful representations of tissue morphology, the generator will reconstruct the held attributes in the latent vectors.
    
    We use the encoder to find the latent vector of real tissue images, and then use the generator on those same vector representations to obtain the tissue image reconstruction. Figure \ref{fig:real_recon} shows these reconstruction results. Although the reconstruction does not have a one-to-one match at the pixel level, we are judging our model by how it finds high level features and assigns representations based on them. The reconstructions keep the same tissue attributes whether if we analyze stromal, lymphocytes, or cancer cells. 
    
    \begin{figure}[!b]
        \centering
        \includegraphics[scale=0.065]{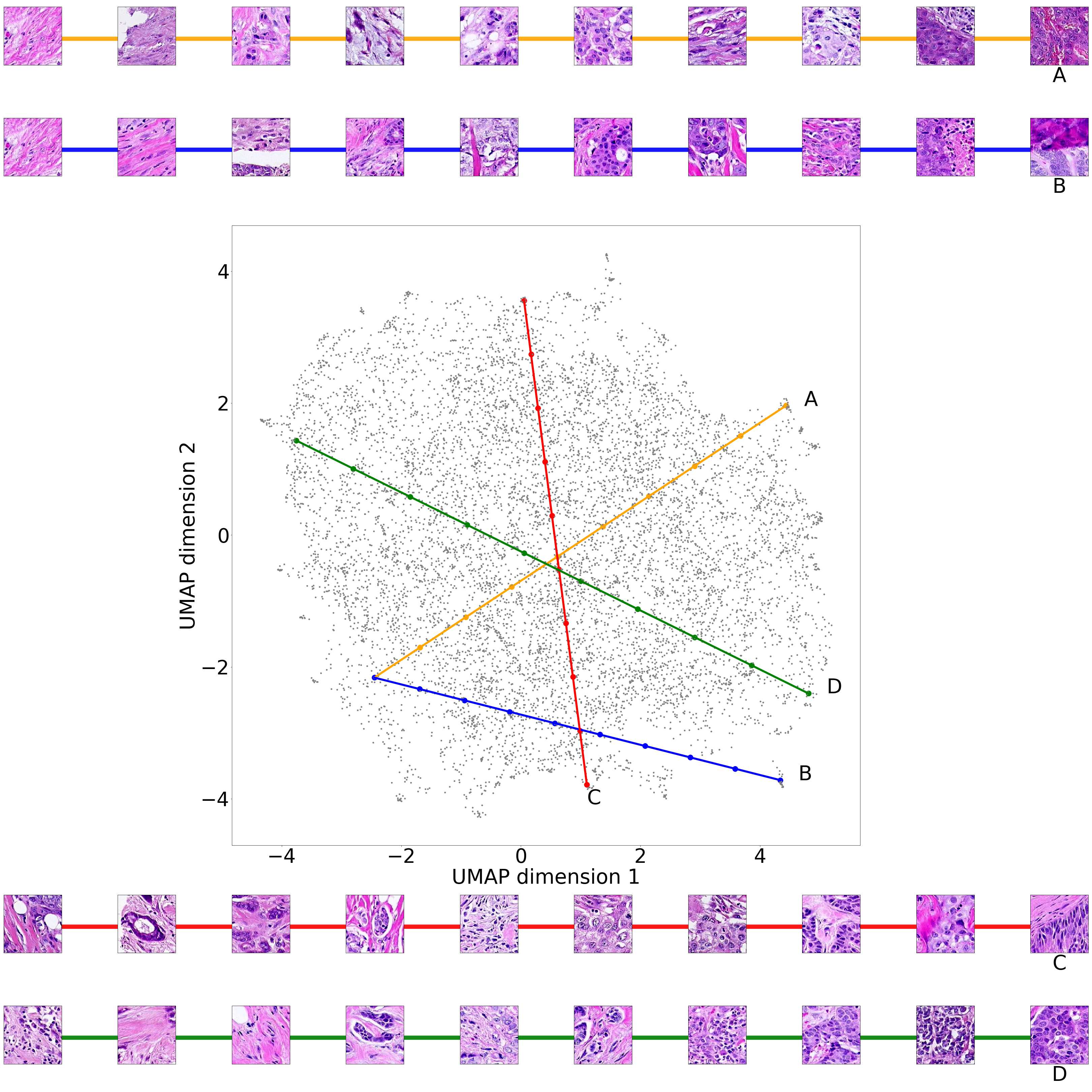}
        \caption{Four different linear
        interpolations between clusters in extreme positions of the latent space at ten equally distributed points each. In contrast to Figure \ref{fig:real_img_cluster}, this figure shows the global structure of the latent space where consecutive image points have gradual morphological changes in the tissue.}
        \label{fig:real_img_linear_interp}
    \end{figure}
    
\subsection{Analysis of real tissue representations}
    We also study the latent space representations of all available real tissue images of VGH and NKI cohorts. We perform a Uniform Manifold Approximation and Projection (UMAP) \cite{lel2018umap} reduction on the $w$ latent vectors from $200$ to $2$ dimensions, and then fit a Gaussian mixture model of $100$ components to cluster the tissue points. We reason that a good representation should have the following properties: 1) points in close proximity should encode similar tissue architectures; 2) far apart points should encode drastically different tissue architectures; 3) changes in tissue architectures should correspond to smooth manifolds in the representation space.   
    
    Figure \ref{fig:real_img_cluster} shows the UMAP plot, which demonstrates that tissue points belonging to the same cluster have common characteristics, not only color and texture,  but also cell types presented in the tissue. Figure \ref{fig:real_img_linear_interp} captures the global structure of the latent space, by displaying four different linear interpolations between clusters across extreme positions of the latent space. Each interpolation is made of ten equally distributed points. We can see that transitions between consecutive points show morphological similarities without abrupt changes. Between Figures \ref{fig:real_img_cluster} and \ref{fig:real_img_linear_interp}, we conclude that our model learns to capture and place gradual changes in cancer tissue, where images with similar morphology can be clustered together. 
    
\subsection{Analysis of survival data using latent representations}
    \begin{figure}[!tb]
        \centering
        \includegraphics[scale=0.075]{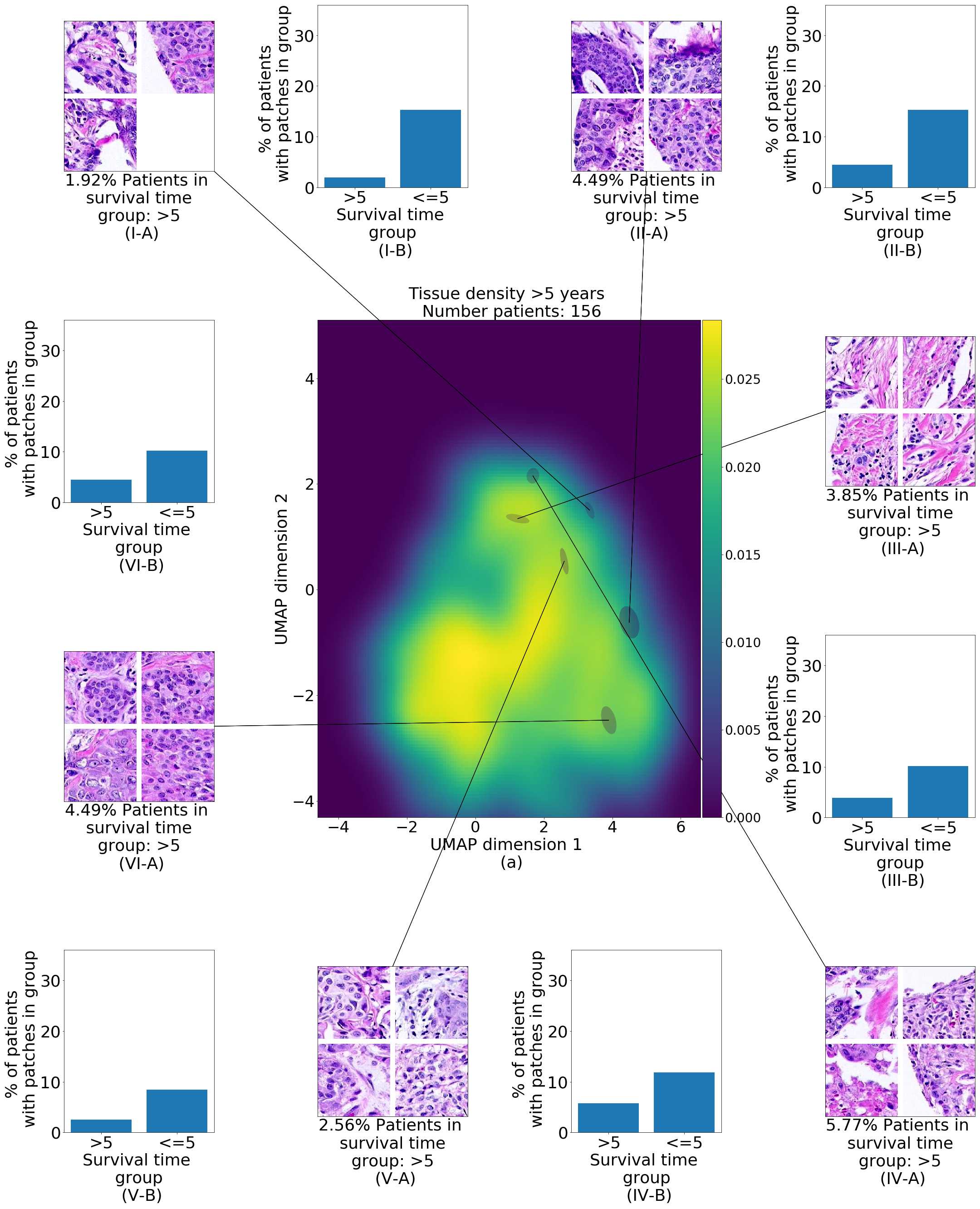}
        \includegraphics[scale=0.075]{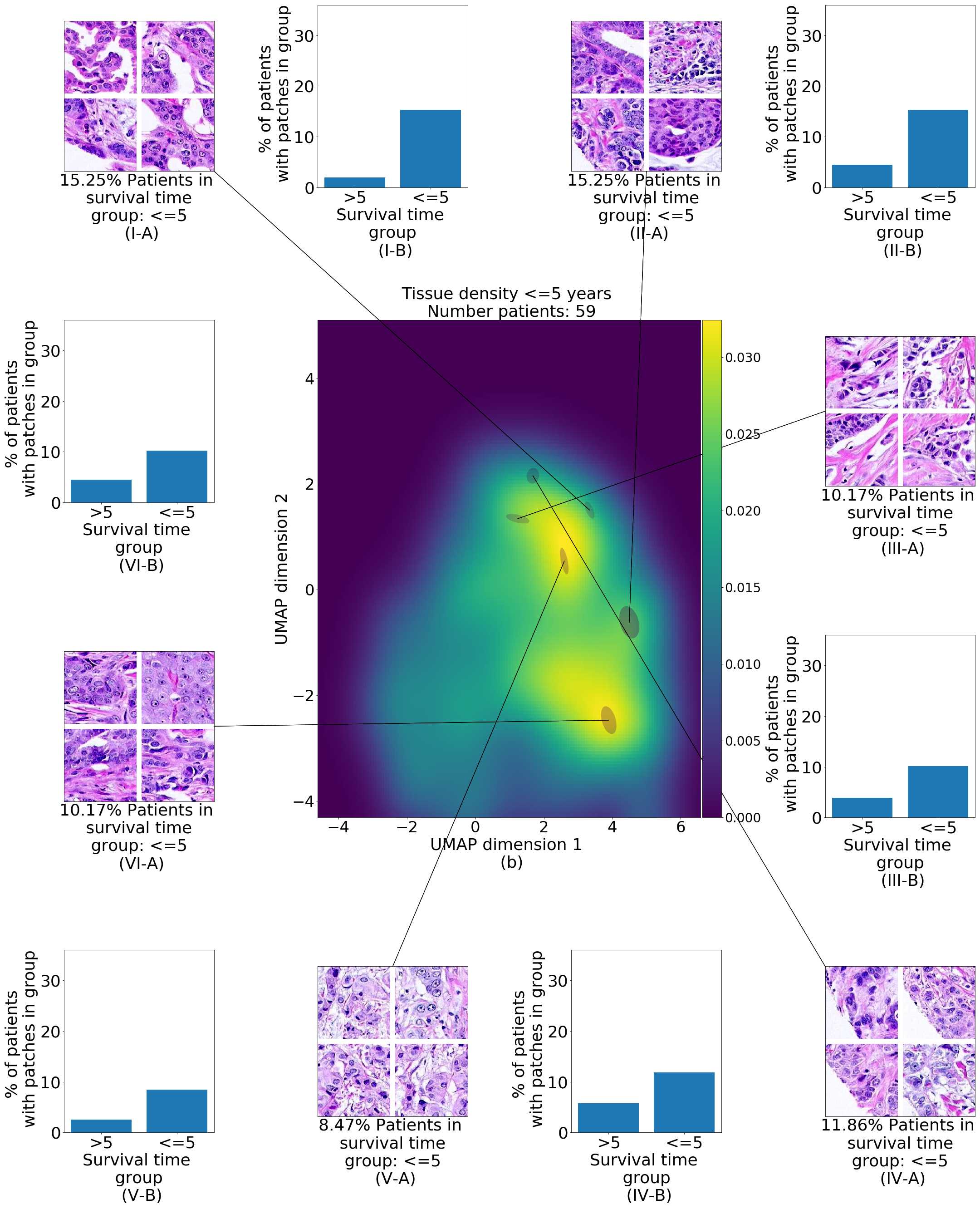}
        \caption{Densities of tissue architectures in patients with greater (a) and lesser (b) than 5 year survival of VGH cohort. We highlight six Gaussian mixture components of tissue architectures that are enriched in high-risk (less than 5 year survival) patients. (\#-A) Tissue images of in the cluster, (\#-B) Percentage of patients with the tissue pattern in the survival group.}
        \label{fig:survival_real_gen}
    \end{figure}
    
    The NKI cohort consists of $198$ patients with survival times greater than five years, and $49$ patients with survival times lesser or equal to five years. In the case of VGH, $156$, $59$ with survival times greater, and lesser or equal to 5 years respectively. We use the clustering properties over tissue morphology of our model, to show the differences in tissue density between patients with greater and lesser than 5 years of survival time. Previous literature found different tissue architectures can dictate improved or worse prognosis \cite{beck/systematic_analysis/2011}. 
    
    
    Figure \ref{fig:survival_real_gen} highlights tissue architectures that are enriched in patients with lesser than 5 year survival, but less frequent in cases with greater than 5 year survival. Across both VGH and NKI cohorts, we find a wide spread distinctly enriched clusters of tissue architectures. These results are included in the Appendix.
    

\section{Conclusion and future work}
    We presented an improvement to PathologyGAN that includes an encoder and can integrate real tissue image data. We showed that this model distinguishes features of real tissue, such as color, texture, cancer, lymphocyte, and stromal cells; the model assigns low dimensional representations that maintain meaning associated with its morphological characteristics. Furthermore, it revealed distinctly enriched clusters of tissue architectures in the high-risk patient groups.
    
    This model opens the door for identifying common/distinct patterns of tissue architecture. This could greatly improve our understanding of tumor mircoenvironment and its relation to patient outcome and underlying molecular characteristics. We are working towards generalizing our findings across large patient cohorts such as The Cancer Genome Atlas.   
    
\bibliographystyle{splncs04}
\bibliography{samplepaper}

\newpage
\appendix
\section{Code}
\label{appendix:code}
    We provide the code at this  \href{https://github.com/AdalbertoCq/Learning-a-low-dimensional-manifold-of-realcancer-tissue}{location}.

\section{Tissue image Reconstruction}
\label{appendix:recon}
    Real tissue images and its reconstructions. We take real tissue images and map them to the latent space with our encoder, then we use the generator with the latent vector representations to generate the image reconstructions. 
    
    In all these samples we use the following labeling: (a) correspond to the real tissue images and (b) to the reconstructions, the images are paired in columns. We show different examples of stromal, lymphocytes, cancer cells, and combinations of these, the reconstructions follow the real image attributes.
    

    \begin{figure}[H]
        \centering
        \includegraphics[scale=0.142]{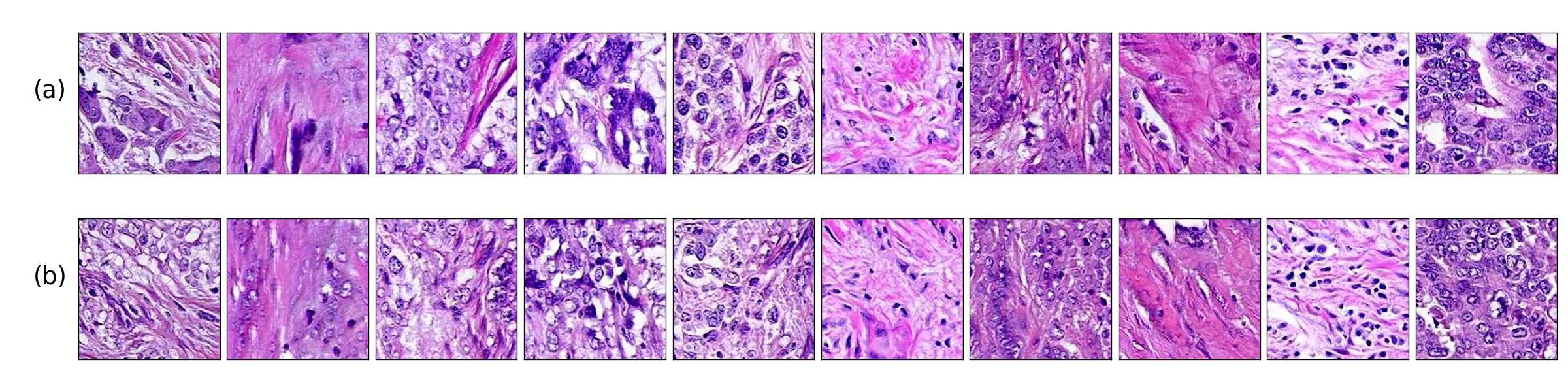}
        \includegraphics[scale=0.142]{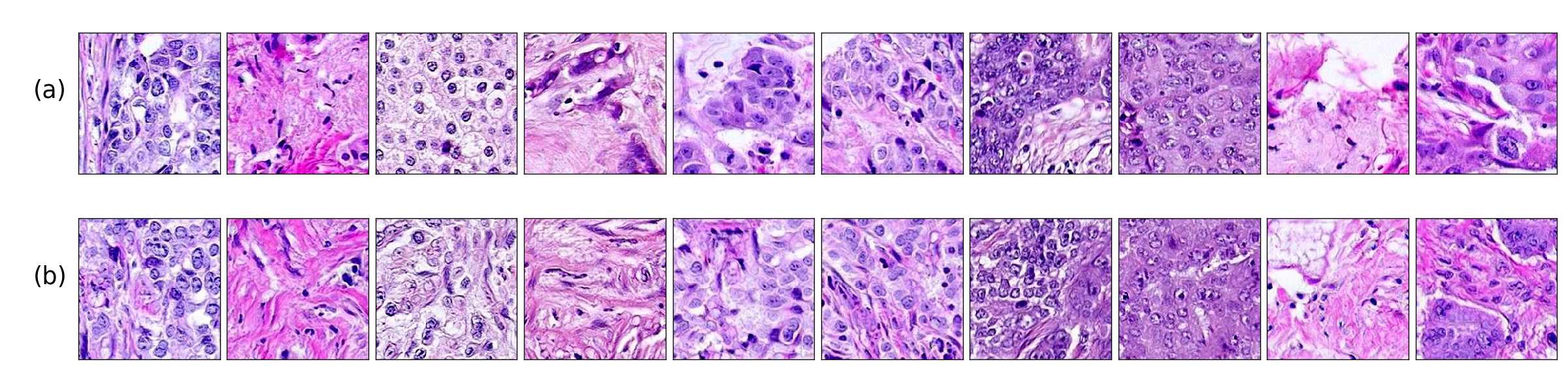}
        \includegraphics[scale=0.142]{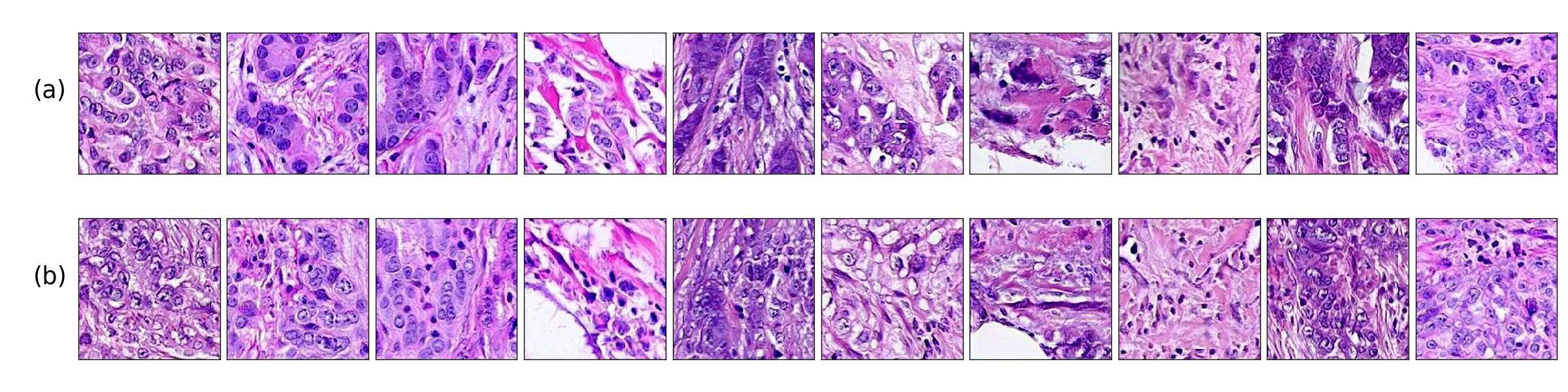}
        \caption{Real tissue images and its reconstructions. (a) corresponds to the real tissue images and (b) to the reconstructions, the images are paired in columns. We show different examples of stromal, lymphocytes, cancer cells, and combinations of these, the reconstructions follow the real image attributes.}
        \label{fig:appendix_real_recon_1}
    \end{figure}
    
    \begin{figure}[H]
        \centering
        \includegraphics[scale=0.142]{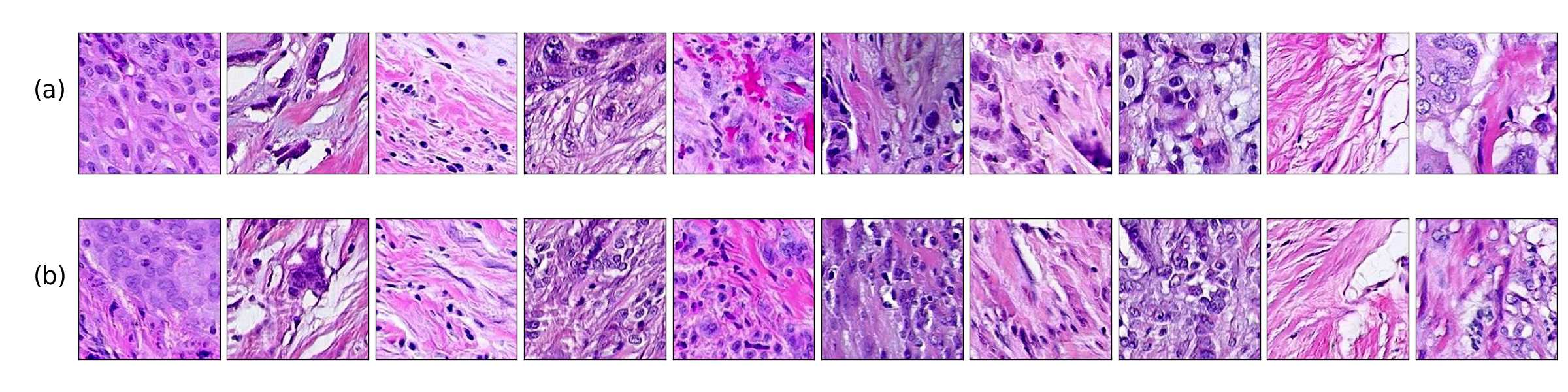}
        \includegraphics[scale=0.142]{images/results/real_recon/Real_reconstructed_PathologyGAN_Enc_Incr_4.jpg}
        \includegraphics[scale=0.142]{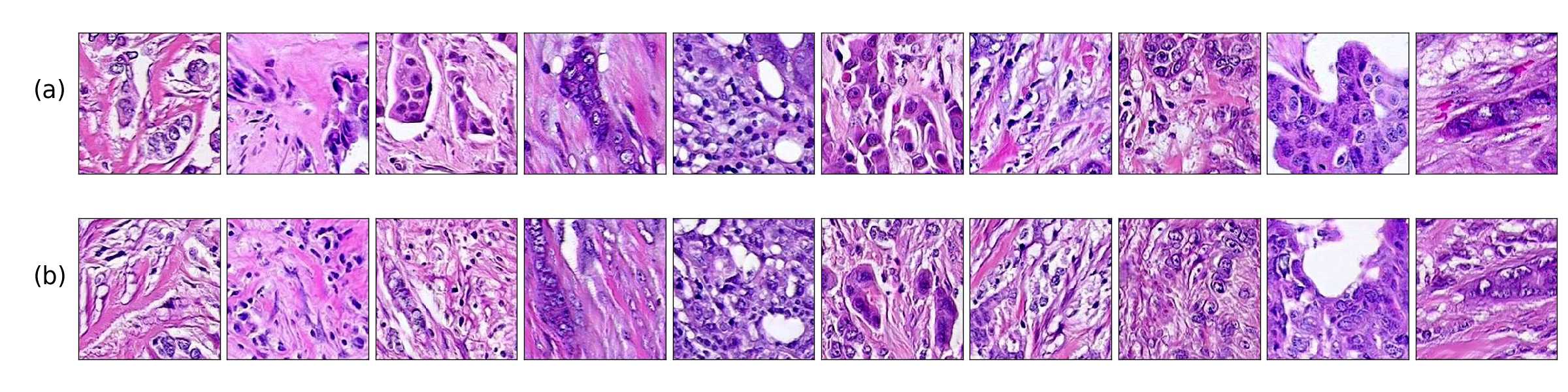}
        \caption{Real tissue images and its reconstructions. (a) corresponds to the real tissue images and (b) to the reconstructions, the images are paired in columns. We show different examples of stromal, lymphocytes, cancer cells, and combinations of these, the reconstructions follow the real image attributes.}
        \label{fig:appendix_real_recon_2}
    \end{figure}
    
\section{Analysis of real tissue representations}
\label{appendix:real_rep}
    Uniform Manifold Approximation and Projection (UMAP) representation of real tissue samples in our model's latent space using samples from Netherlands Cancer Institute (NKI) and Vancouver General Hospital (VGH) patient cohorts. In this Figure, we placed a Gaussian mixture model over the complete dataset and used 100 components to cluster the latent representations. 
    
\begin{figure}[H]
        \centering
        \includegraphics[scale=0.19]{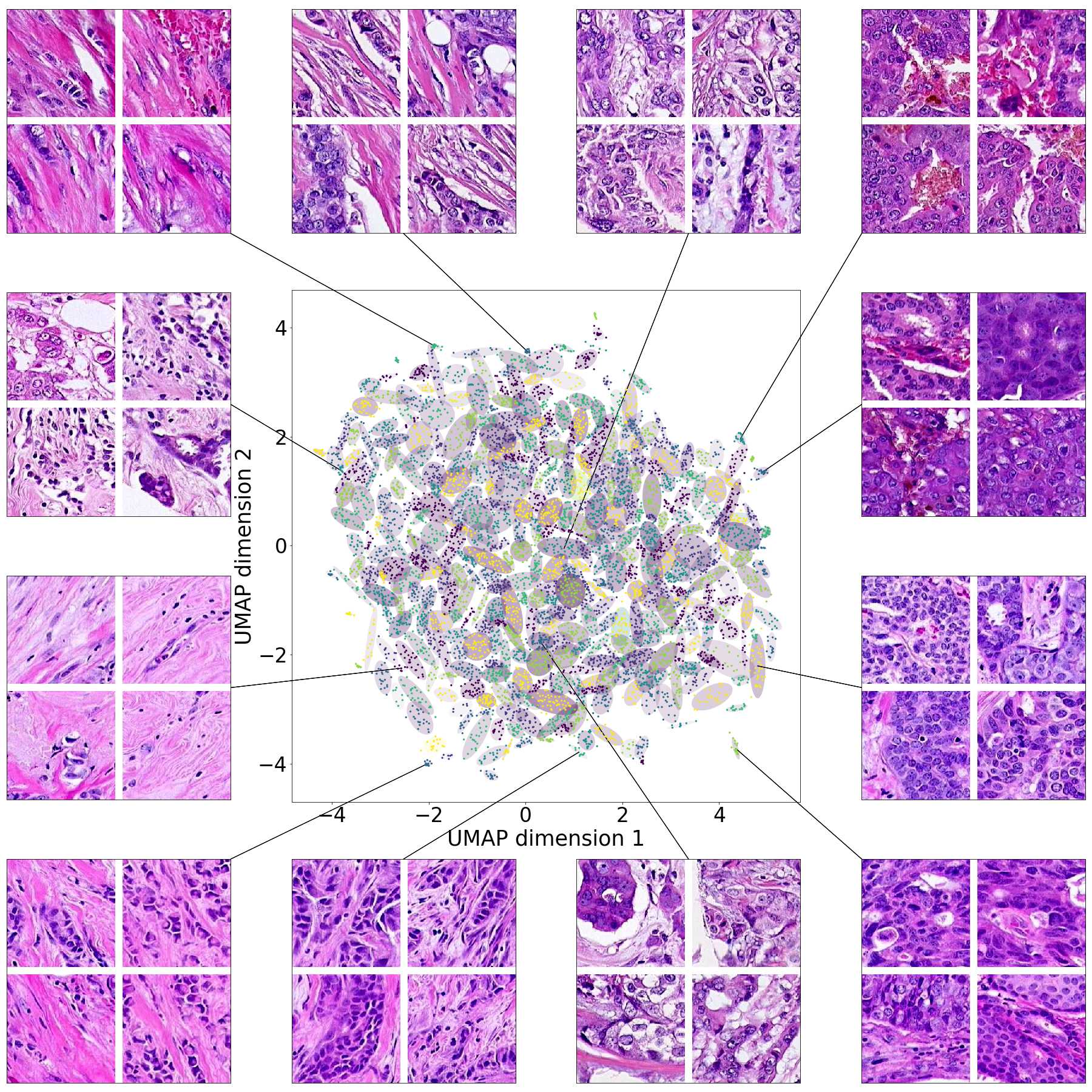}
        \caption{Uniform Manifold Approximation and Projection (UMAP) representation of real tissue samples in the latent space using samples from Netherlands Cancer Institute (NKI) and Vancouver General Hospital (VGH) patient cohorts. In this Figure, we fitted a Gaussian mixture model over the complete dataset and used 100 components to cluster the latent representations. We show different tissue images belonging to various unique clusters, demonstrating how tissues with similar features get assigned to common regions in the latent space.}
        \label{fig:appendix_real_img_cluster_VGH_NKI}
    \end{figure}

\begin{figure}[H]
        \centering
        \includegraphics[scale=0.11]{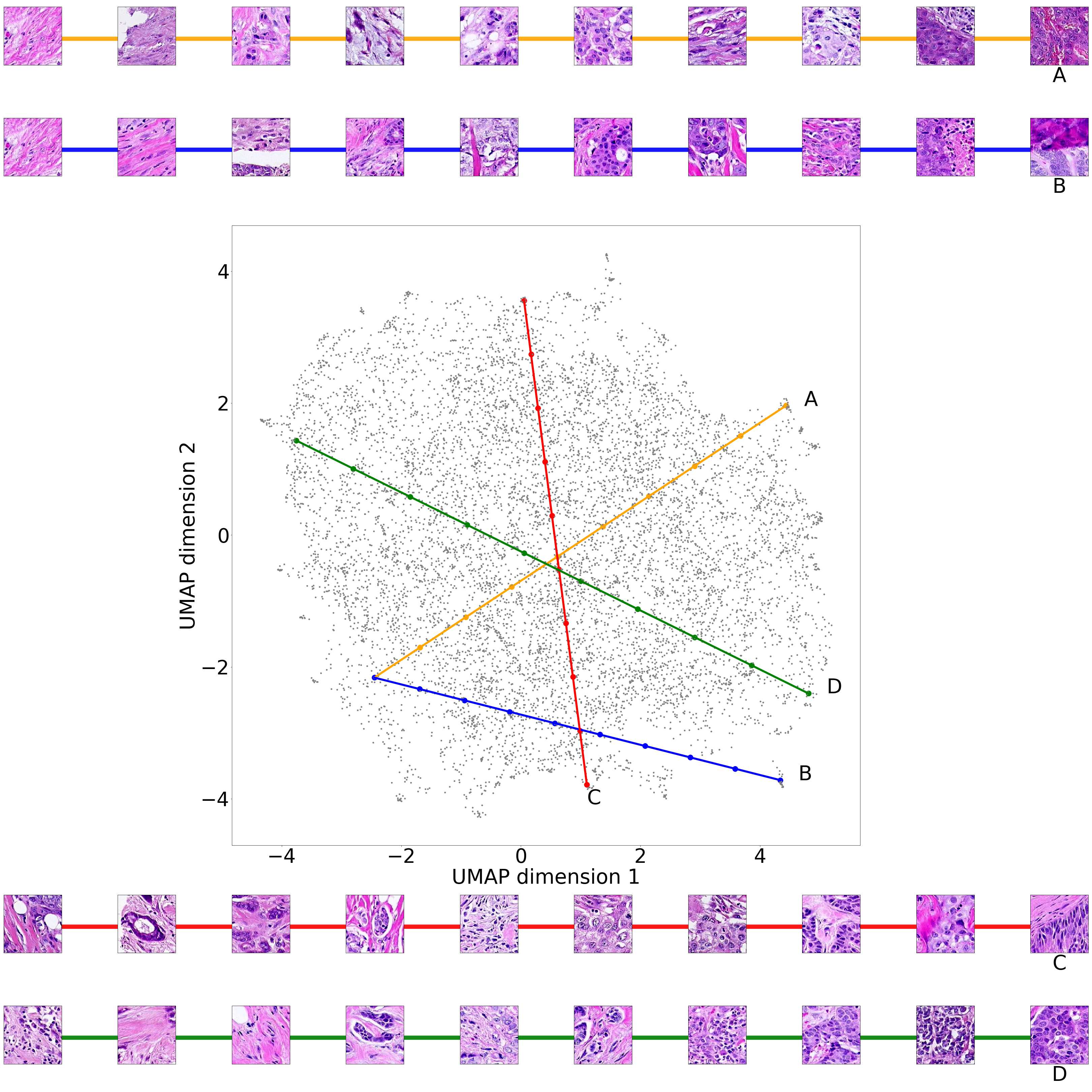}
        \caption{Four different linear
        interpolations between clusters in extreme positions of the latent space at ten equally distributed points each. In contrast to Figure \ref{fig:appendix_real_img_cluster_VGH_NKI}, this figure shows the global structure of the latent space where consecutive image points have gradual morphological changes in the tissue.}
        \label{fig:appendix_real_img_linear_interp_VGH_NKI}
    \end{figure}

\begin{figure}[H]
        \centering
        \includegraphics[scale=0.19]{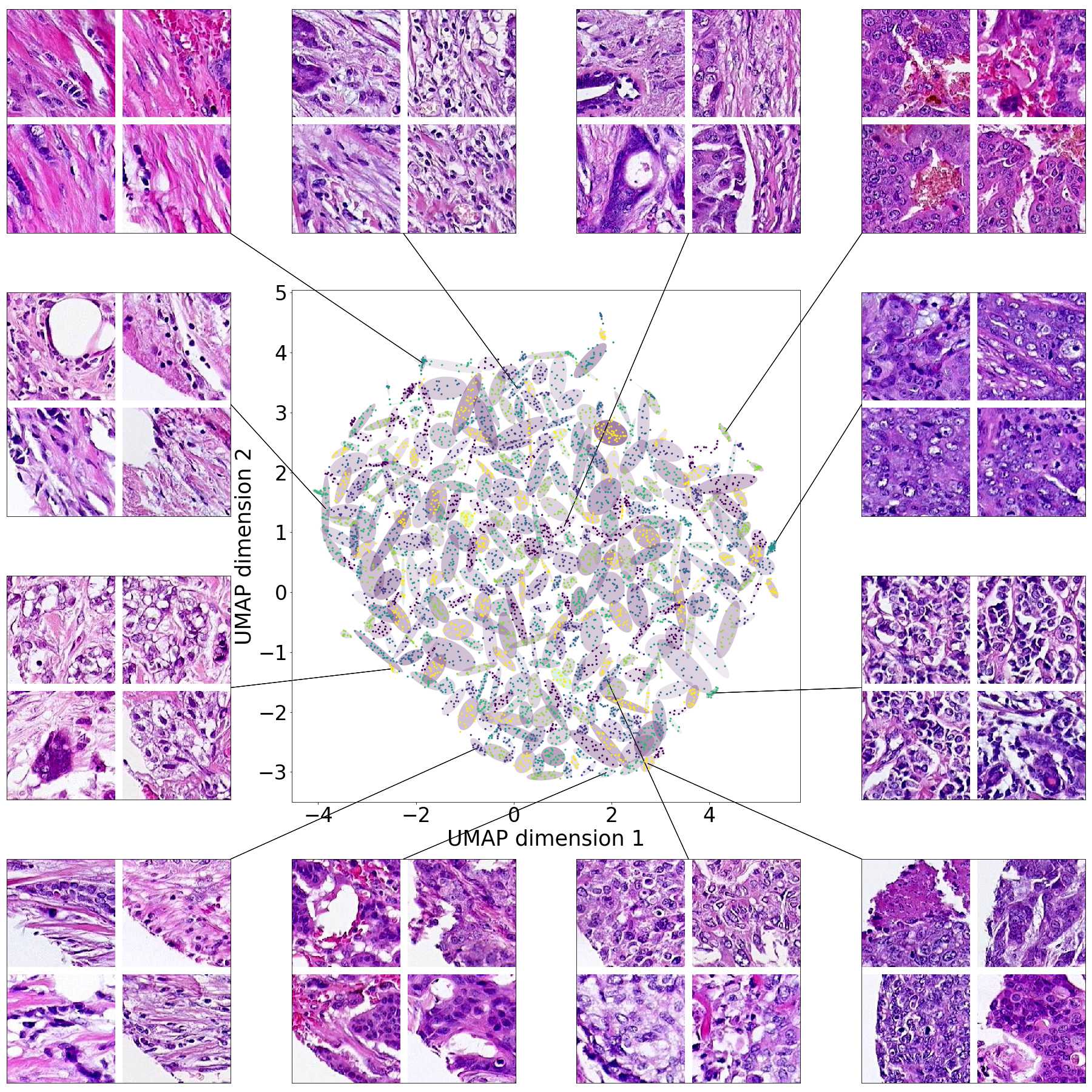}
        \caption{Uniform Manifold Approximation and Projection (UMAP) representation of real tissue samples in the latent space using samples from Netherlands Cancer Institute (NKI) patient cohorts. In this Figure, we fitted a Gaussian mixture model over the complete dataset and used 100 components to cluster the latent representations. We show different tissue images belonging to various unique clusters, demonstrating how tissues with similar features get assigned to common regions in the latent space.}
        \label{fig:appendix_real_img_cluster_NKI}
    \end{figure}

\begin{figure}[H]
        \centering
        \includegraphics[scale=0.11]{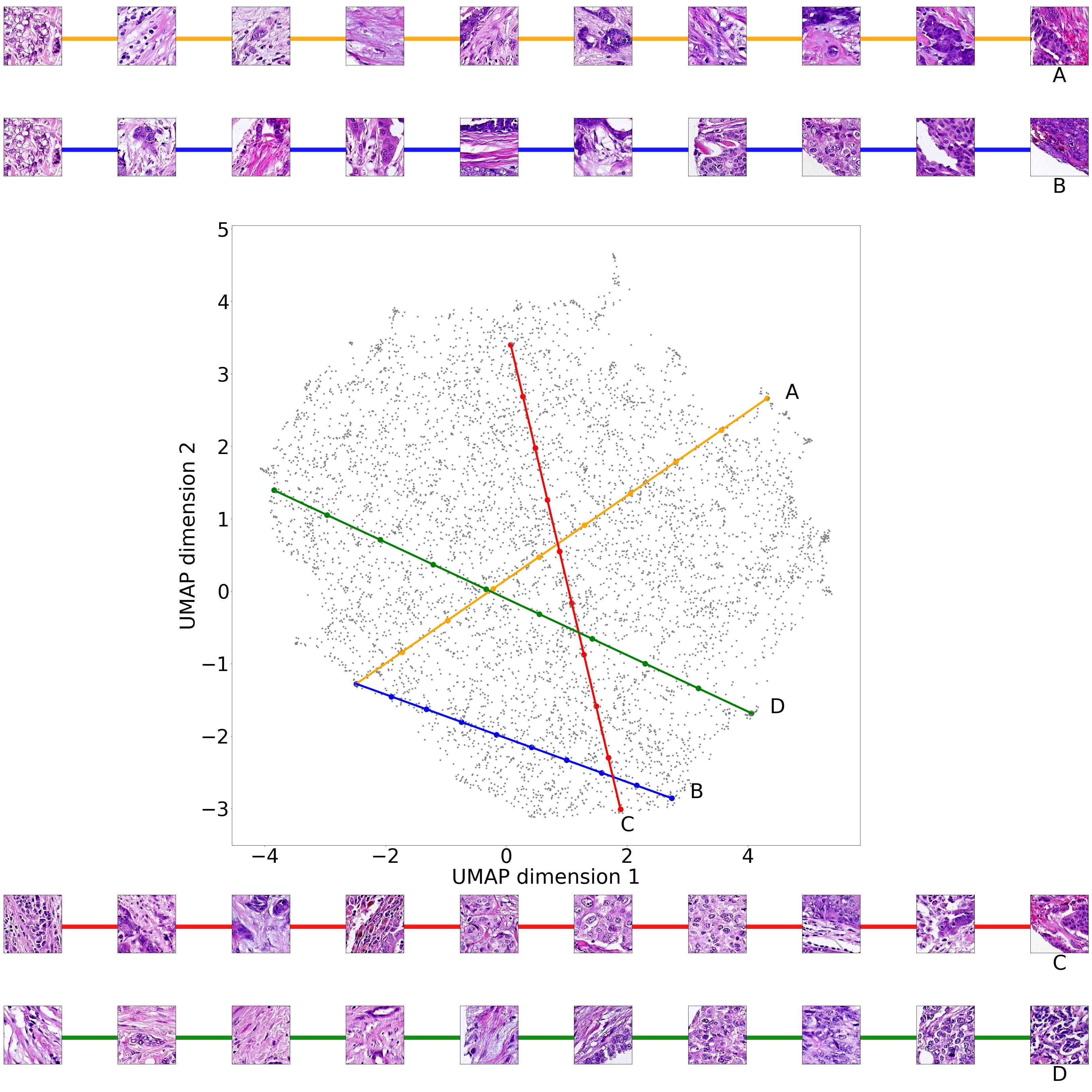}
        \caption{Four different linear
        interpolations between clusters in extreme positions of the latent space at ten equally distributed points each. In contrast to Figure \ref{fig:appendix_real_img_cluster_NKI}, this figure shows the global structure of the latent space where consecutive image points have gradual morphological changes in the tissue in the NKI patient cohort.}
        \label{fig:appendix_real_img_linear_interp_NKI}
    \end{figure}

\begin{figure}[H]
        \centering
        \includegraphics[scale=0.19]{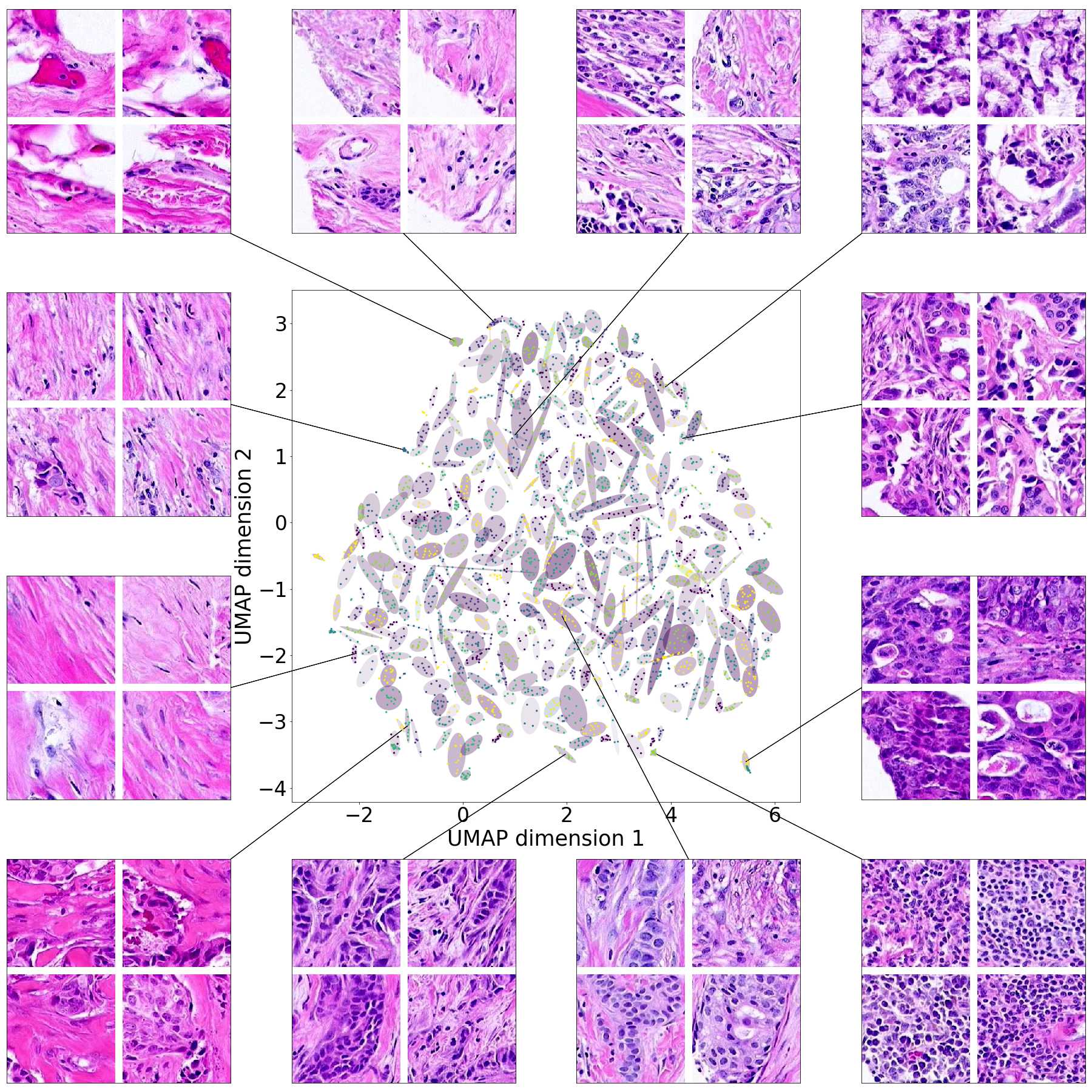}
        \caption{Uniform Manifold Approximation and Projection (UMAP) representation of real tissue samples in the latent space using samples from Vancouver General Hospital (VGH) patient cohorts. In this Figure, we fitted a Gaussian mixture model over the complete dataset and used 100 components to cluster the latent representations. We show different tissue images belonging to various unique clusters, demonstrating how tissues with similar features get assigned to common regions in the latent space.}
        \label{fig:appendix_real_img_cluster_VGH}
    \end{figure}

\begin{figure}[H]
        \centering
        \includegraphics[scale=0.11]{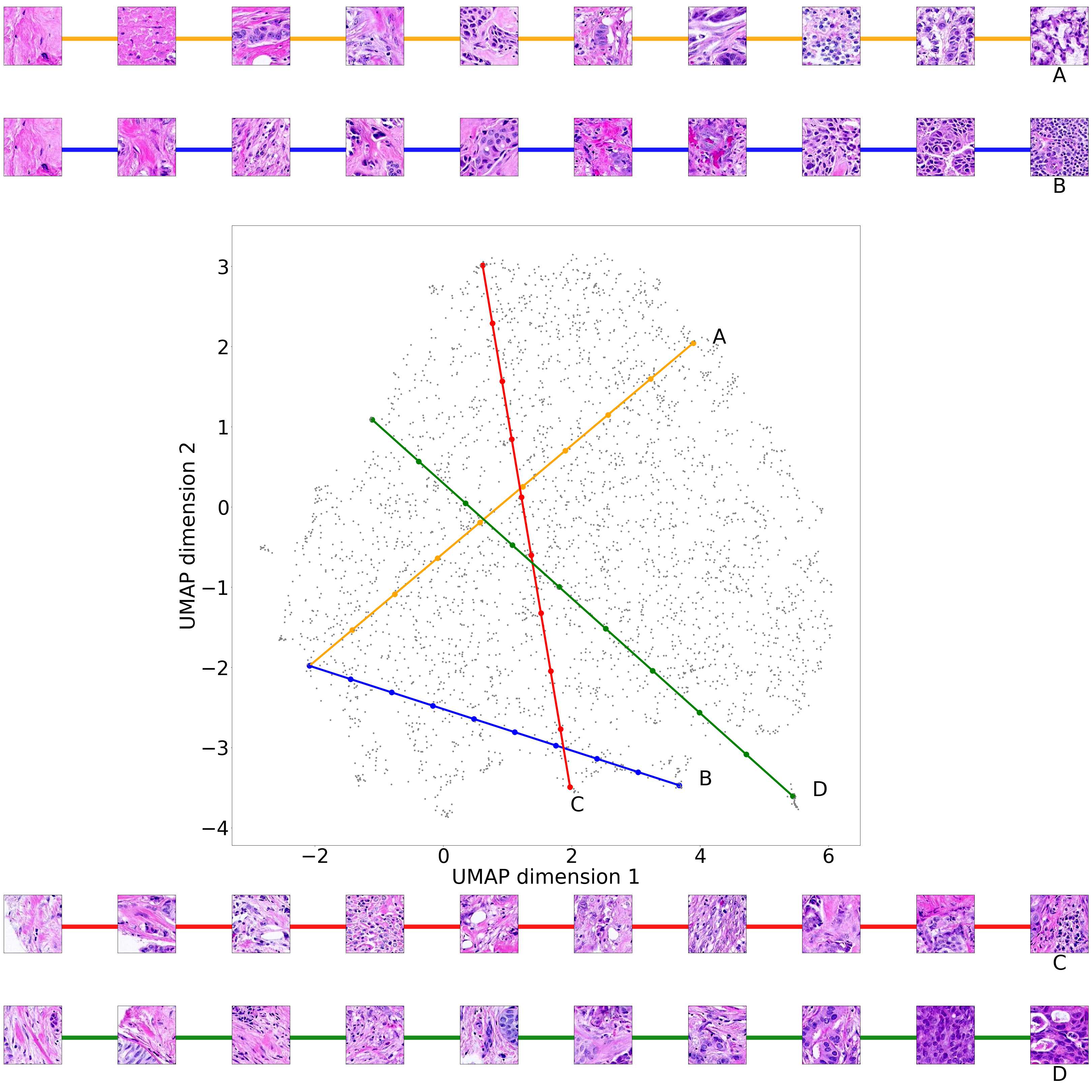}
        \caption{Four different linear
        interpolations between clusters in extreme positions of the latent space at ten equally distributed points each. In contrast to Figure \ref{fig:appendix_real_img_cluster_VGH}, this figure shows the global structure of the latent space where consecutive image points have gradual morphological changes in the tissue in the VGH patient cohort.}
        \label{fig:appendix_real_img_linear_interp_VGH}
    \end{figure}

    We show two different type of figures for the combined datasets VGH and NKi, and for NKI and VGH independently:
    \begin{enumerate}
        \item Clustering of tissue architectures into common regions of the latent space, Figures \ref{fig:appendix_real_img_cluster_VGH_NKI},\ref{fig:appendix_real_img_cluster_NKI},\ref{fig:appendix_real_img_cluster_VGH}.
        \item Global structure of the latent space, Figures \ref{fig:appendix_real_img_linear_interp_VGH_NKI},\ref{fig:appendix_real_img_linear_interp_NKI},\ref{fig:appendix_real_img_linear_interp_VGH}.
    \end{enumerate}
    
\section{Analysis of survival data using latent representations}
\label{appendix:real_survival}
    In this appendix we provide the collection of figures for NKI and VGH patient cohorts. We show the density difference in tissue architectures between high-risk (less than 5 year survival (a)) and low-risk patients (greater than 5 year survival (b)). 
    
\begin{figure}[H]
        \centering
        \includegraphics[scale=0.07]{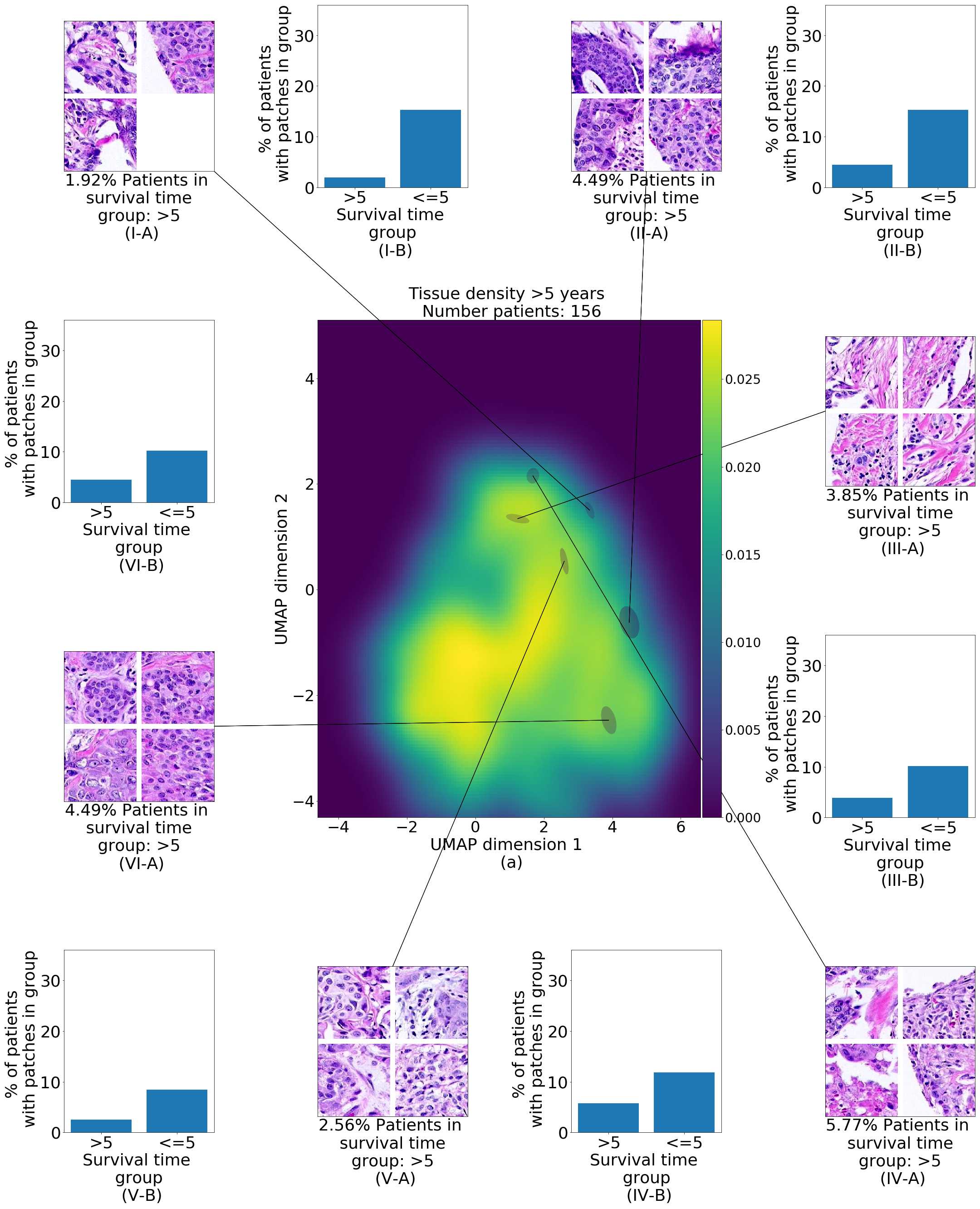}
        \includegraphics[scale=0.07]{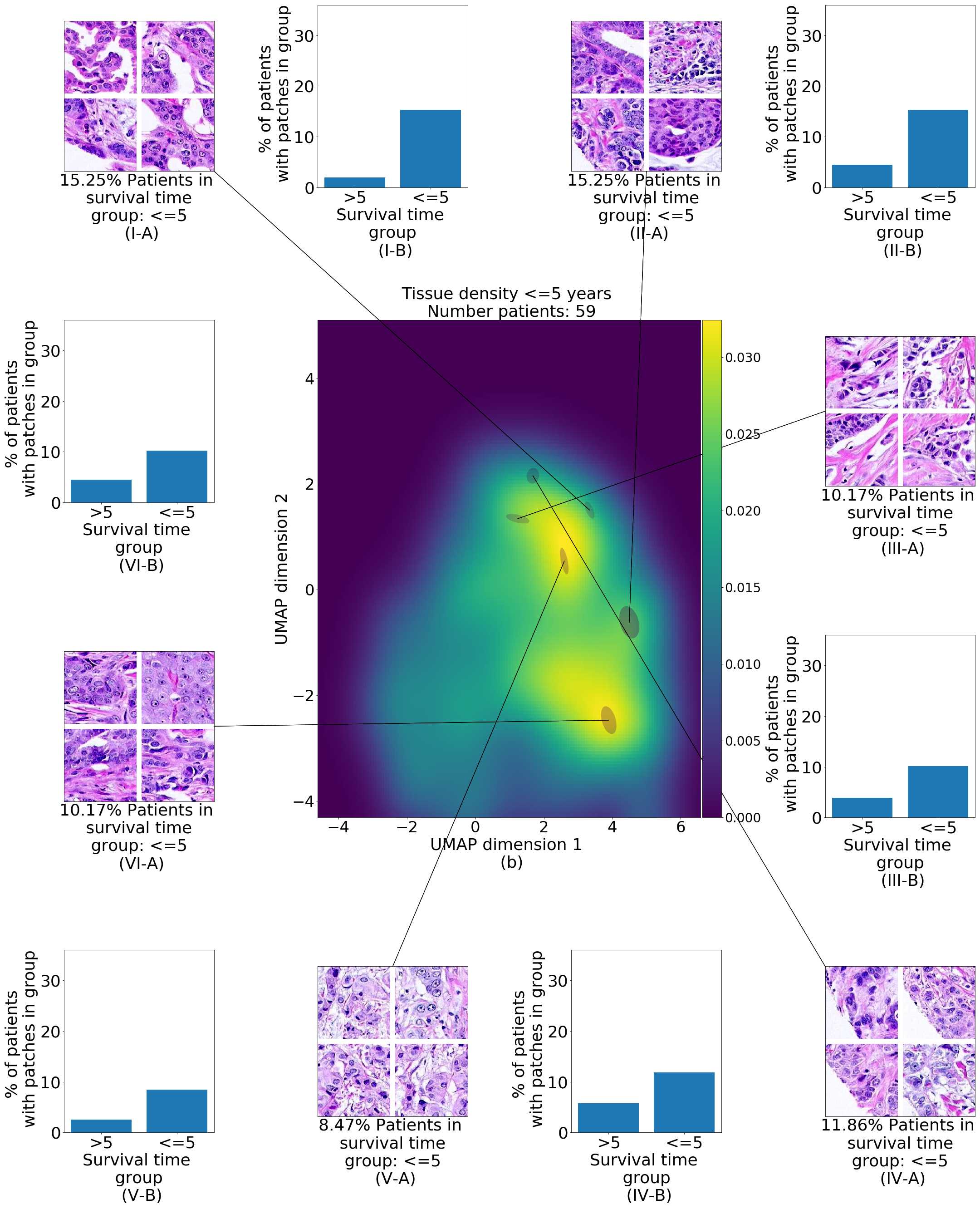}
        \caption{VGH Cohort, tissue architecture more predominant on high-risk patients, survival times lesser than 5 years.}
        \label{fig:appendix_high_risk_VGH}
    \end{figure}

\begin{figure}[H]
        \centering
        \includegraphics[scale=0.075]{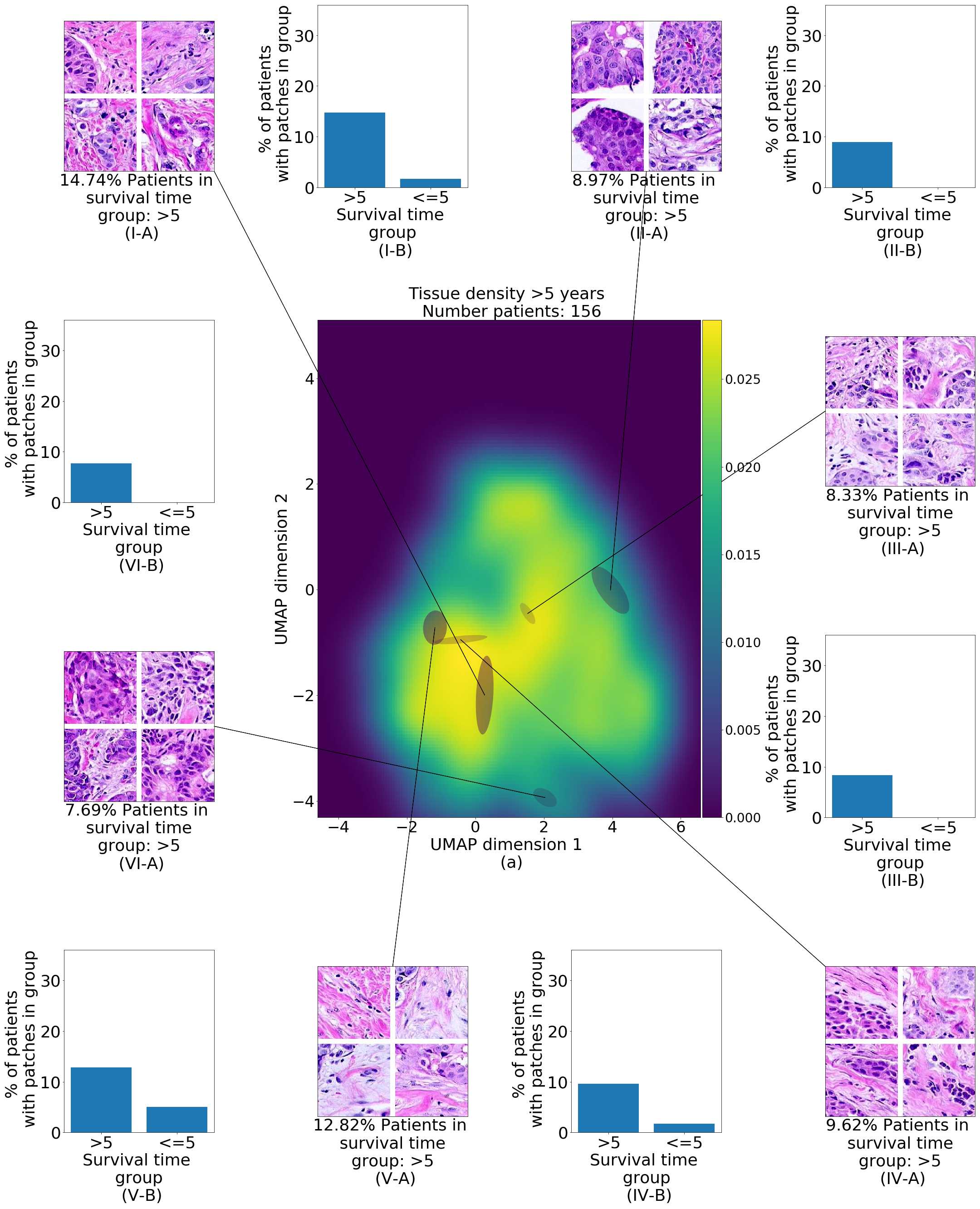}
        \includegraphics[scale=0.075]{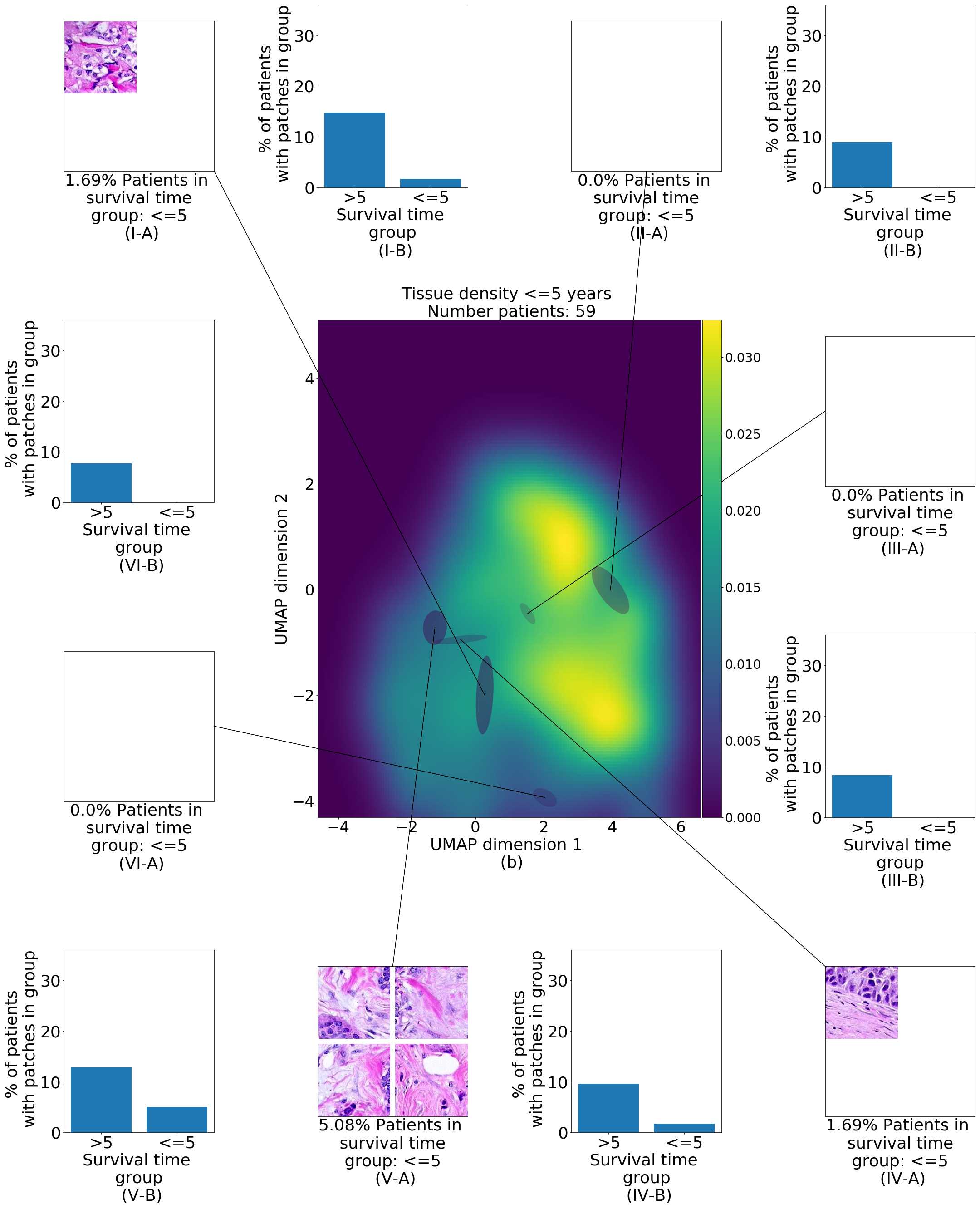}
        \caption{VGH Cohort, tissue architecture more predominant on low-risk patients, survival times greater than 5 years.}
        \label{fig:appendix_low_risk_VGH}
    \end{figure}

\begin{figure}[H]
        \centering
        \includegraphics[scale=0.07]{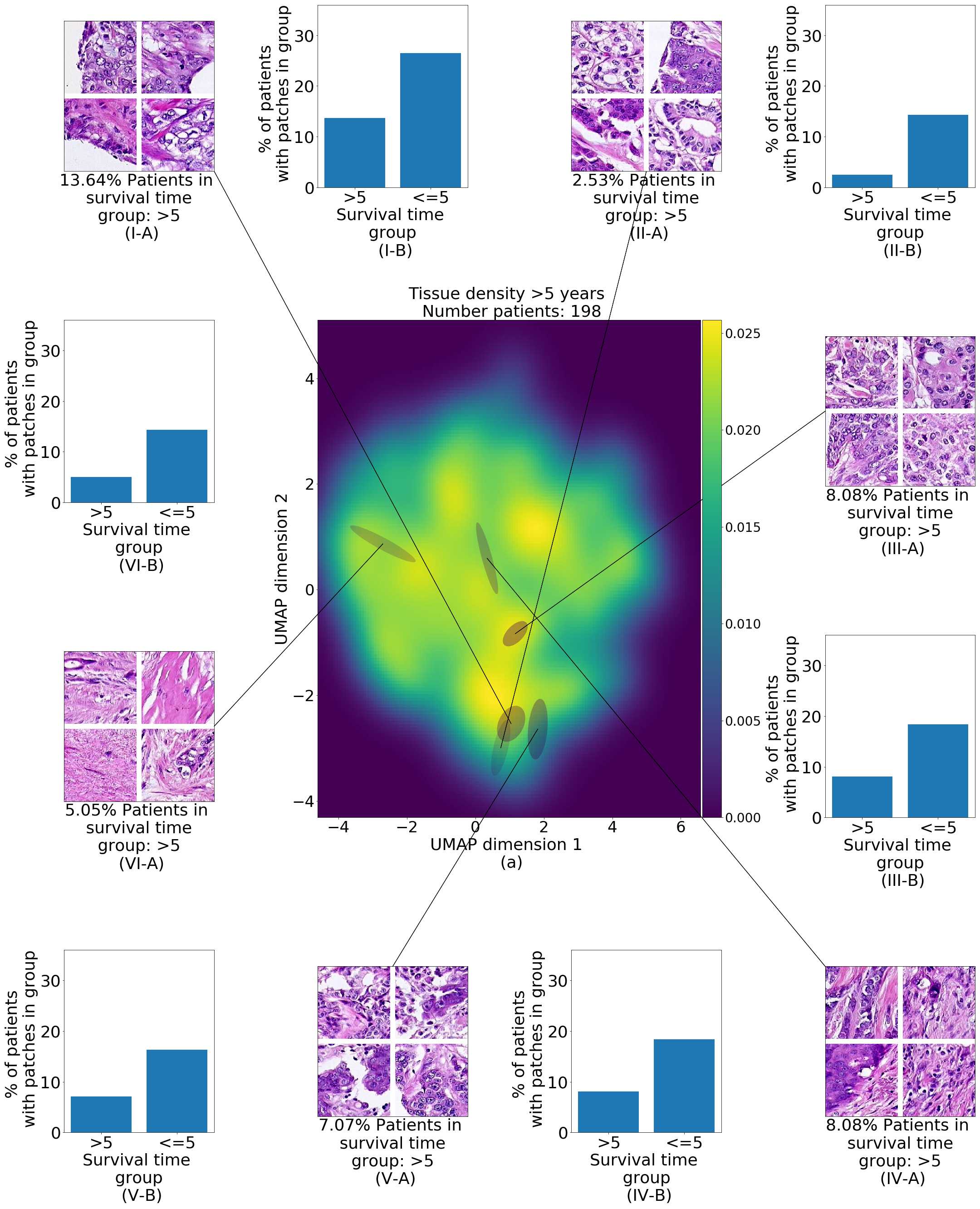}
        \includegraphics[scale=0.07]{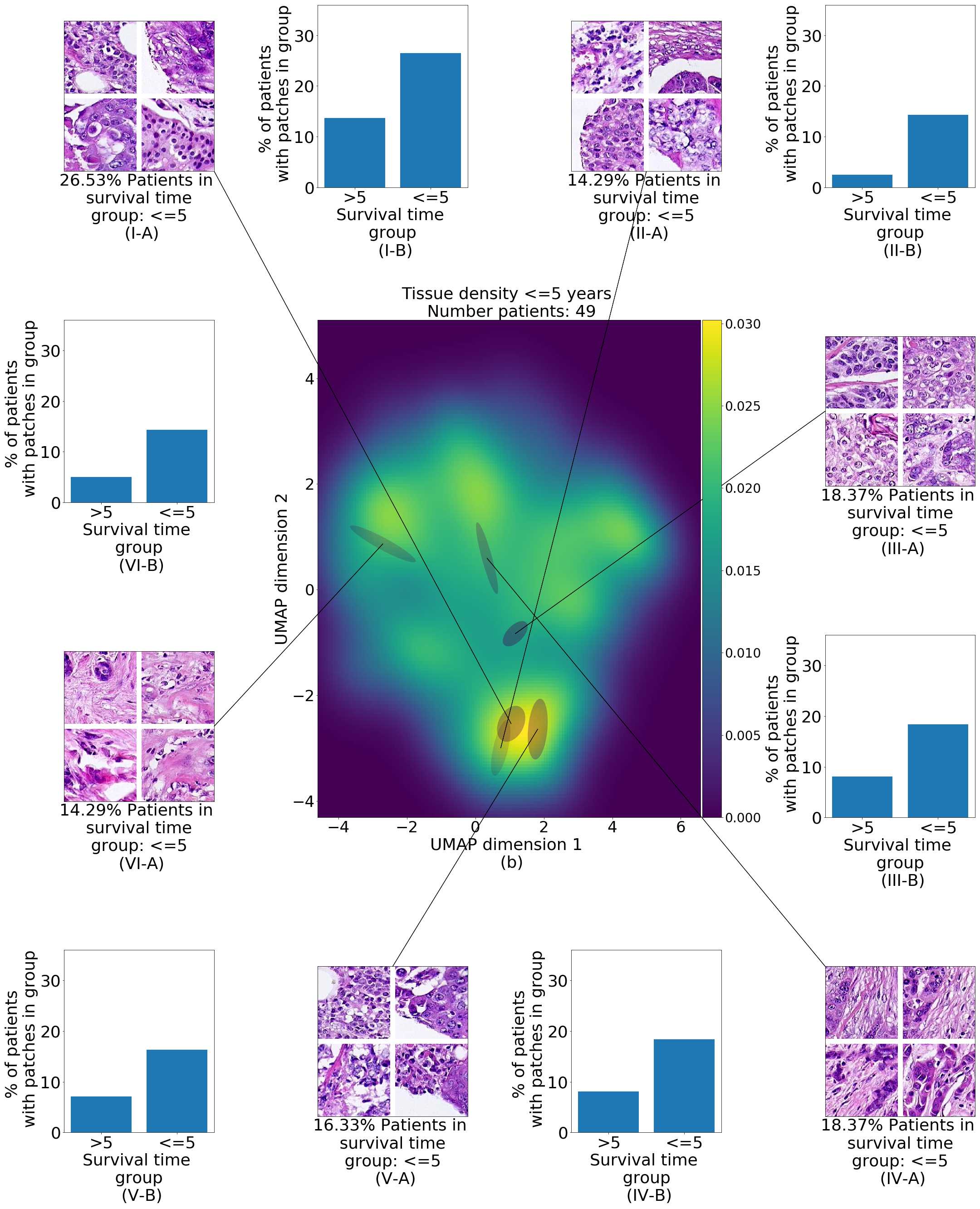}
        \caption{NKI Cohort, tissue architecture more predominant on high-risk patients, survival times lesser than 5 years.}
        \label{fig:appendix_high_risk_NKI}
    \end{figure}

\begin{figure}[H]
        \centering
        \includegraphics[scale=0.07]{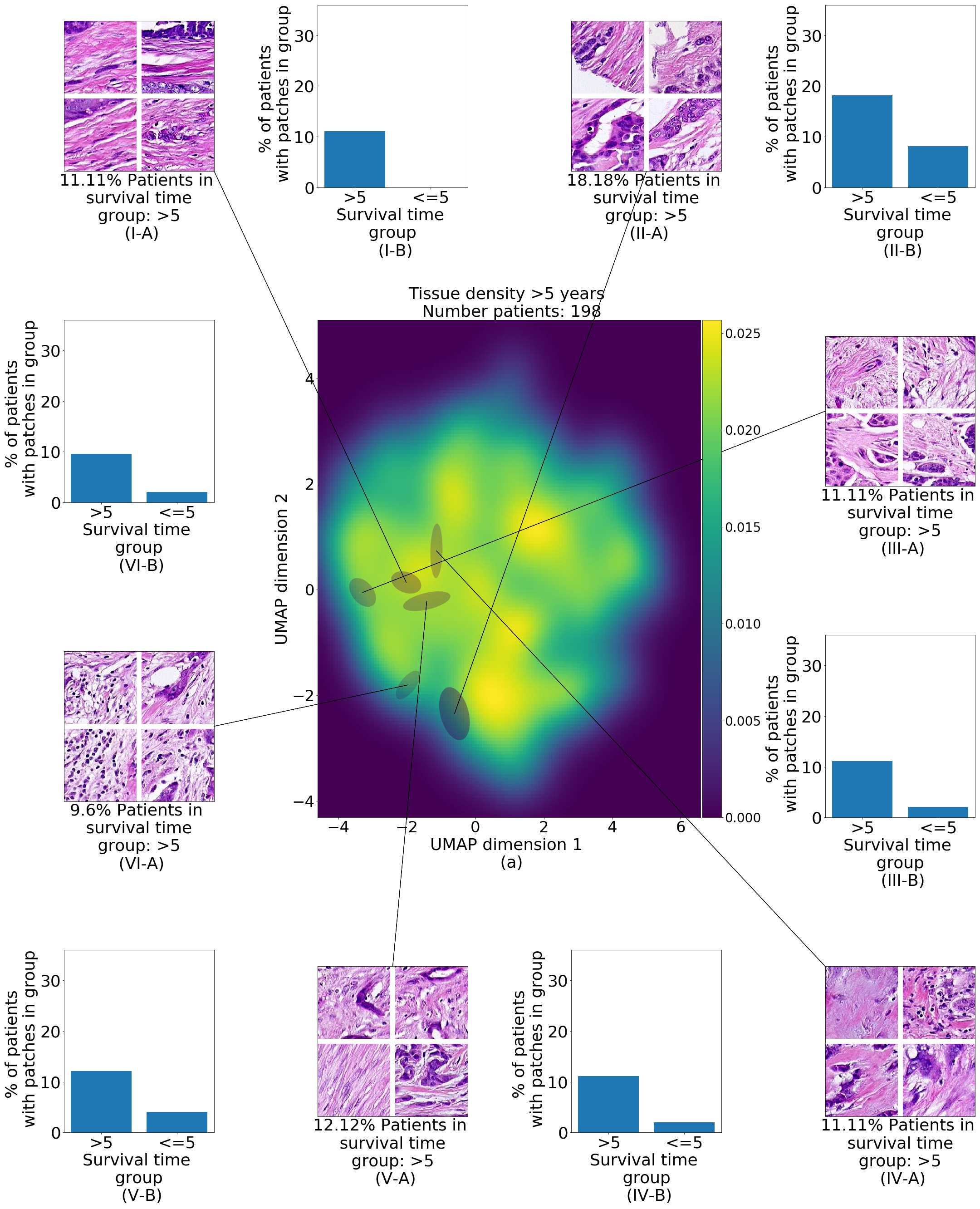}
        \includegraphics[scale=0.07]{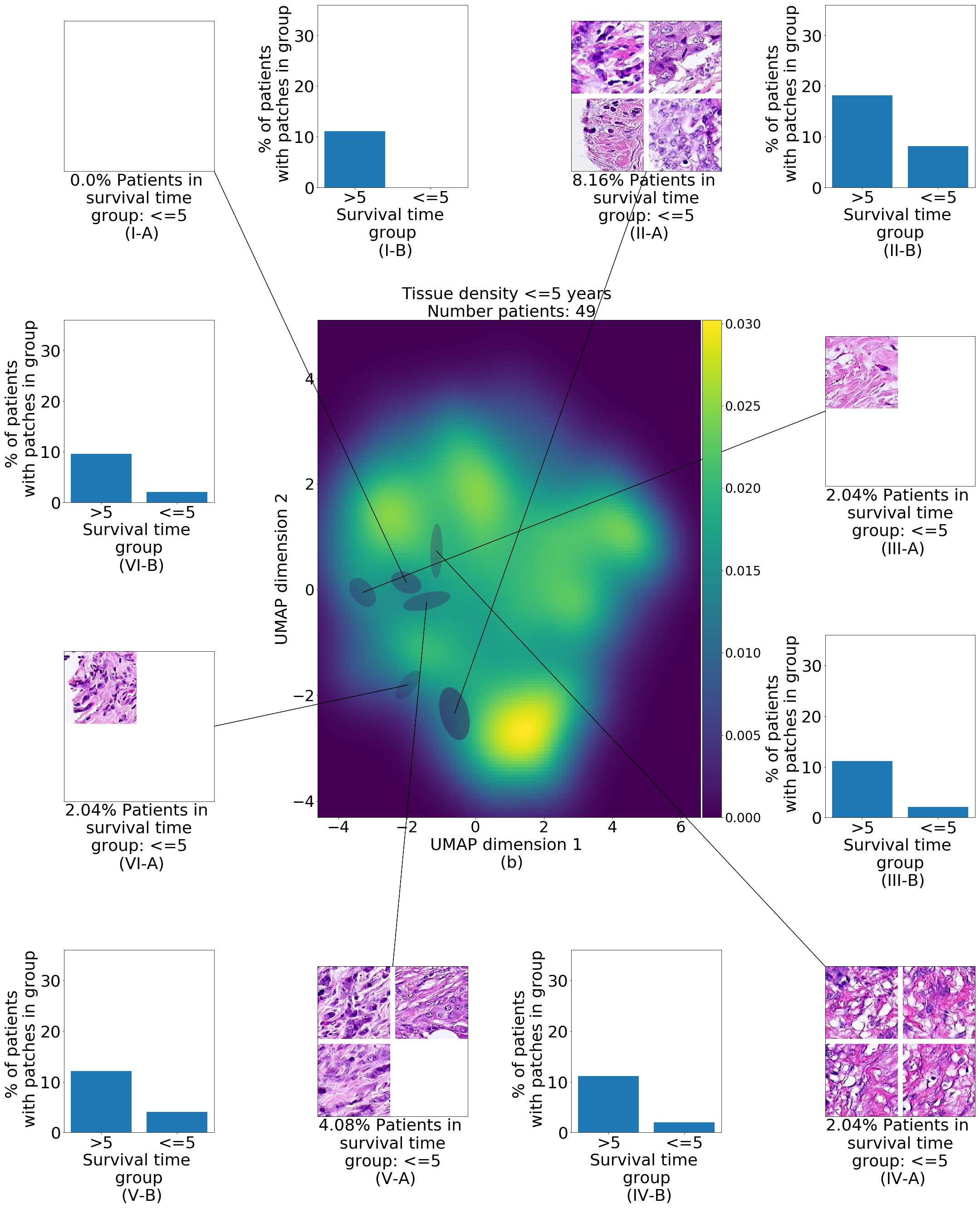}
        \caption{NKI Cohort, tissue architecture more predominant on low-risk patients, survival times greater than 5 years.}
        \label{fig:appendix_low_risk_NKI}
    \end{figure}

    Figures \ref{fig:appendix_high_risk_VGH} and \ref{fig:appendix_high_risk_NKI} present tissue architectures predominant on high-risk patients, and Figures \ref{fig:appendix_low_risk_VGH} and \ref{fig:appendix_low_risk_NKI} on low-risk patients. 
    
    (\#-A) Tissue images belonging to the cluster, (\#-B) Percentage of patients with the tissue pattern in the survival group.
    
    

\section{Model Architecture}
\label{appendix:model_architecture}
    
    \begin{table}[H]
        \centering
        \begin{tabular}{c}
        Mapping Network $M:z \rightarrow w$ \\
        \toprule
        \midrule
        $z \in  \sim \mathbb{R}^{200} \sim \mathcal{N}(0, I)$ \\
        \midrule
        ResNet Dense Layer and ReLU, $200 \rightarrow 200$ \\
        \midrule
        ResNet Dense Layer and ReLU, $200 \rightarrow 200$ \\
        \midrule
        ResNet Dense Layer and ReLU, $200 \rightarrow 200$ \\
        \midrule
        ResNet Dense Layer and ReLU, $200 \rightarrow 200$ \\
        \midrule
        Dense Layer, $200 \rightarrow 200$ \\
        \bottomrule
        \bottomrule
        \end{tabular}
        \vspace*{2mm}
        \caption{Mapping Network Architecture details of Pathology GAN model.}
        \label{mapping_network_arch}
    \end{table}
    
    \newpage
    
    \begin{table}[H]
        \centering
        \begin{tabular}{c}
        Generator Network $G:w \rightarrow x$ \\
        \toprule
        \midrule
        Dense Layer, adaptive instance normalization (AdaIN), and leakyReLU \\
        $200 \rightarrow 1024$ \\
        \midrule
        Dense Layer, AdaIN, and leakyReLU \\
        $1024 \rightarrow 12544$ \\
        \midrule
        Reshape $7\times7\times256$ \\
        \midrule
        ResNet Conv2D Layer, 3x3, stride 1, pad same, AdaIN, and leakyReLU $0.2$ \\
        $ 7\times7\times256 \rightarrow 7\times7\times256 $ \\
        \midrule
        ConvTranspose2D Layer, 2x2, stride 2, pad upscale, AdaIN, and leakyReLU $0.2$ \\
        $ 7\times7\times256 \rightarrow 14\times14\times512 $ \\
        \midrule
        ResNet Conv2D Layer, 3x3, stride 1, pad same, AdaIN, and leakyReLU $0.2$ \\
        $ 14\times14\times512 \rightarrow 14\times14\times512 $ \\
        \midrule
        ConvTranspose2D Layer, 2x2, stride 2, pad upscale, AdaIN, and leakyReLU $0.2$ \\
        $ 14\times14\times512 \rightarrow 28\times28\times256 $ \\
        \midrule
        ResNet Conv2D Layer, 3x3, stride 1, pad same, AdaIN, and leakyReLU $0.2$ \\
        $ 28\times28\times256 \rightarrow 28\times28\times256 $ \\
        \midrule
        Attention Layer at $28\times28\times256$ \\
        \midrule
        ConvTranspose2D Layer, 2x2, stride 2, pad upscale, AdaIN, and leakyReLU $0.2$ \\
        $ 28\times28\times256 \rightarrow 56\times56\times128 $ \\
        \midrule
        ResNet Conv2D Layer, 3x3, stride 1, pad same, AdaIN, and leakyReLU $0.2$ \\
        $ 56\times56\times128 \rightarrow 56\times56\times128 $\\
        \midrule
        ConvTranspose2D Layer, 2x2, stride 2, pad upscale, AdaIN, and leakyReLU $0.2$ \\
        $ 56\times56\times128 \rightarrow 112\times112\times64 $ \\
        \midrule
        ResNet Conv2D Layer, 3x3, stride 1, pad same, AdaIN, and leakyReLU $0.2$ \\
        $ 112\times112\times64 \rightarrow 112\times112\times64 $ \\
        \midrule
        ConvTranspose2D Layer, 2x2, stride 2, pad upscale, AdaIN, and leakyReLU $0.2$ \\
        $ 112\times112\times64 \rightarrow 224\times224\times32 $ \\
        \midrule
        Conv2D Layer, 3x3, stride 1, pad same, $ 32 \rightarrow 3 $ \\
        $ 224\times224\times32 \rightarrow 224\times224\times3 $ \\
        \midrule
        Sigmoid \\
        \bottomrule
        \bottomrule
        \end{tabular}
        \vspace*{2mm}
        \caption{Generator Network Architecture details of Pathology GAN model.}
        \label{generator_arch}
    \end{table}
    
    \begin{table}[H]
        \centering
        \begin{tabular}{c}
        Discriminator Network $C:x \rightarrow d$ \\
        \toprule
        \midrule
        $x \in  \mathbb{R}^{224\times224\times3}$ \\
        \midrule
        ResNet Conv2D Layer, 3x3, stride 1, pad same, and leakyReLU $0.2$ \\
        $ 224\times224\times3 \rightarrow 224\times224\times3 $ \\
        \midrule
        Conv2D Layer, 2x2, stride 2, pad downscale, and leakyReLU $0.2$ \\
        $ 224\times224\times3 \rightarrow 122\times122\times32 $ \\
        \midrule
        ResNet Conv2D Layer, 3x3, stride 1, pad same, and leakyReLU $0.2$ \\
        $ 122\times122\times32 \rightarrow 122\times122\times32 $ \\
        \midrule
        Conv2D Layer, 2x2, stride 2, pad downscale, and leakyReLU $0.2$ \\
        $ 122\times122\times32 \rightarrow 56\times56\times64 $ \\
        \midrule
        ResNet Conv2D Layer, 3x3, stride 1, pad same, and leakyReLU $0.2$ \\
        $ 56\times56\times64 \rightarrow 56\times56\times64 $ \\
        \midrule
        Conv2D Layer, 2x2, stride 2, pad downscale, and leakyReLU $0.2$ \\
        $ 56\times56\times64 \rightarrow 28\times28\times128 $ \\
        \midrule
        ResNet Conv2D Layer, 3x3, stride 1, pad same, and leakyReLU $0.2$ \\
        $ 28\times28\times128 \rightarrow 28\times28\times128 $ \\
        \midrule
        Attention Layer at $28\times28\times128$ \\
        \midrule
        Conv2D Layer, 2x2, stride 2, pad downscale, and leakyReLU $0.2$ \\
        $ 28\times28\times128 \rightarrow 14\times14\times256 $ \\
        \midrule
        ResNet Conv2D Layer, 3x3, stride 1, pad same, and leakyReLU $0.2$ \\
        $ 14\times14\times256 \rightarrow 14\times14\times256 $ \\
        \midrule
        Conv2D Layer, 2x2, stride 2, pad downscale, and leakyReLU $0.2$ \\
        $ 14\times14\times256 \rightarrow 7\times7\times512 $ \\
        \midrule
        Flatten $ 7\times7\times512 \rightarrow 25088 $ \\
        \midrule
        Dense Layer and leakyReLU, $25088 \rightarrow 1024$ \\
        \midrule
        Dense Layer and leakyReLU, $1024 \rightarrow 1$ \\
        \bottomrule
        \bottomrule
        \end{tabular}
        \vspace*{2mm}
        \caption{Discriminator Network Architecture details of Pathology GAN model.}
        \label{Discriminator_arch}
    \end{table}
    
    \begin{table}[H]
        \centering
        \begin{tabular}{c}
        Encoder Network $E:x \rightarrow w'$ \\
        \toprule
        \midrule
        $x \in  \mathbb{R}^{224\times224\times3}$ \\
        \midrule
        Conv2D Layer, 2x2, stride 2, pad downscale, and leakyReLU $0.2$ \\
        $ 224\times224\times3 \rightarrow 224\times224\times32 $ \\
        \midrule
        ResNet Conv2D Layer, 3x3, stride 1, pad same, and leakyReLU $0.2$ \\
        $ 224\times224\times32 \rightarrow 224\times224\times32 $ \\
        \midrule
        Conv2D Layer, 2x2, stride 2, pad downscale, and leakyReLU $0.2$ \\
        $ 224\times224\times32 \rightarrow 122\times122\times64 $ \\
        \midrule
        ResNet Conv2D Layer, 3x3, stride 1, pad same, and leakyReLU $0.2$ \\
        $ 122\times122\times64 \rightarrow 122\times122\times64 $ \\
        \midrule
        Conv2D Layer, 2x2, stride 2, pad downscale, and leakyReLU $0.2$ \\
        $ 122\times122\times64 \rightarrow 56\times56\times128 $ \\
        \midrule
        ResNet Conv2D Layer, 3x3, stride 1, pad same, and leakyReLU $0.2$ \\
        $ 56\times56\times128 \rightarrow 56\times56\times128 $ \\
        \midrule
        Conv2D Layer, 2x2, stride 2, pad downscale, and leakyReLU $0.2$ \\
        $ 56\times56\times128 \rightarrow 28\times28\times256 $ \\
        \midrule
        ResNet Conv2D Layer, 3x3, stride 1, pad same, and leakyReLU $0.2$ \\
        $ 28\times28\times256 \rightarrow 28\times28\times256 $ \\
        \midrule
        Attention Layer at $28\times28\times256$ \\
        \midrule
        Conv2D Layer, 2x2, stride 2, pad downscale, and leakyReLU $0.2$ \\
        $ 28\times28\times256 \rightarrow 14\times14\times512 $ \\
        \midrule
        ResNet Conv2D Layer, 3x3, stride 1, pad same, and leakyReLU $0.2$ \\
        $ 14\times14\times512 \rightarrow 14\times14\times512 $ \\
        \midrule
        Conv2D Layer, 2x2, stride 2, pad downscale, and leakyReLU $0.2$ \\
        $ 14\times14\times512 \rightarrow 7\times7\times512 $ \\
        \midrule
        Flatten $ 7\times7\times512 \rightarrow 25088 $ \\
        \midrule
        Dense Layer and leakyReLU, $25088 \rightarrow 1024$ \\
        \midrule
        Dense Layer and leakyReLU, $1024 \rightarrow 200$ \\
        \bottomrule
        \bottomrule
        \end{tabular}
        \vspace*{2mm}
        \caption{Encoder Network Architecture details of Pathology GAN model.}
        \label{Encoder_arch}
    \end{table}

\end{document}